\documentclass{article}

\usepackage{graphicx}
\usepackage{psfig}
\usepackage{epsfig}
\usepackage[round]{natbib}

\title{
Year-ahead Prediction of Hurricane Season Sea Surface Temperature
in the Tropical Atlantic
 }

\begin{document}

\author{Jonathan Meagher\footnote{\emph{Correspondence address}: Email: \texttt{jmeagher@alumni.caltech.edu}},
Stephen Jewson\\}

\maketitle

\begin{abstract}
One possible method for the year-ahead prediction of hurricane numbers would be to
make a year-ahead prediction of sea surface temperature (SST), and then to apply relationships
that link SST to hurricane numbers.
As a first step towards setting up such a system this article
compares three simple statistical methods for the year-ahead prediction of the relevant SSTs.
\end{abstract}

\section{Introduction}

Hurricanes expose the insurance industry to a large amount of risk. This risk varies in space and time
in complex ways, and in order to set insurance premiums at reasonable levels,
it is important to estimate the magnitude and dependencies of
this risk as accurately as possible. One important part of making such an estimate involves predicting the
distribution of possible hurricane behaviour in the future. Since most insurance contracts are a year in length, and
are renewed annually, one of the principal timescales over which hurricane behaviour needs to be predicted
is the annual timescale. This corresponds to what we call `year-ahead prediction' of hurricanes: predicting the
distribution of properties for next year's hurricanes, based on all the information we have at the end of this year's
hurricane season. One particular aspect of the distribution of properties of next year's hurricanes is the
distribution of the \emph{number} of hurricanes, and developing methods for year-ahead predictions of the number of hurricanes
is the topic of this article.

There are a number of methods that one might consider using to try and predict the number of hurricanes a year
in advance.
One set of methods involves searching for statistical predictors of hurricane numbers, and using regression-type
methods to use such predictors to make predictions. This is the method followed by~\citet{klotzbachg04}
and~\citet{saundersl05b}.
Another set of methods is to take the time series of the number of hurricanes per year, study its properties,
and try and make a statistical prediction on that basis. We have investigated two versions of this approach in recent
articles: first, in~\citet{j81} and~\citet{j88}, we have performed back-testing analyses of simple prediction
schemes for the hurricane number time series,
and second, in~\citet{j90}, we
have used shrinkage to combine forecasts based on long and short
baselines. All of these time-series methods derive their prediction skill (if any) from
the trends and long-term variability in the hurricane number
time series.

A third set of methods
would be to consider the underlying causes of any long-term fluctuations in hurricane numbers,
and predict those causes first.
For instance, there is general consensus that much of the
variability in hurricane numbers on long time-scales is related to changes in the ocean, and,
in particular, to changes in the sea surface temperature (SST).
And it is claimed that SST is affected both by long-term cycles
(sometimes known as the Atlantic Multidecadal Oscillation, or AMO: see~\citet{suttonh})
and by climate change, both of which might make it predictable to some extent.
This raises the possibility that one might be able to predict
year-ahead hurricane numbers by first predicting year-ahead sea surface temperatures, and then relating the sea-surface
temperatures to the hurricane numbers. Testing out such a system is our goal.
In this article we will address the question of how to make year-ahead predictions of SST, while
in subsequent articles we intend to address the second half of the problem, which is to relate the SST to
the numbers of hurricanes.


There are many ways that one might consider trying to predict SST. A major division is between
empirical or statistical methods, on the one hand, and physically-based methods on the other.
Empirical methods try to derive statistical relations from the observed historical data.
Physically-based methods attempt to apply the laws of dynamics and thermodynamics using differential equations.
As a first step, we will take a very simple and straightforward empirical approach.
From a methodological point of view, this seems to be the most appropriate way to start addressing this question.
The results from our simple approach can then be used as a benchmark against which other more complex methods
can be judged.

In order to set up an empirical scheme for predicting SST, one has to make some assumptions about
the nature of the variability of SST. This is inevitable: it is not possible to create a prediction scheme
that doesn't make at least one assumption. One could, for instance, assume that the trend in SST is linear,
or that the AMO has a fixed period, and one could derive an optimal prediction scheme for SST, given these
assumptions. However, these particular assumptions are rather restrictive. We will, as an alternative,
make the assumption that the characteristics of predictability of SST are the same now as they have been
in the past. In other words, we assume that whatever methods would have worked well for the prediction of SST in
the past will work well in the future. We then proceed to test and compare a number of simple statistical prediction schemes
on past data, and we conclude that whichever works the best on the past data is the
best scheme to use for our next prediction.
Is this a reasonable approach?
In a changing climate, any assumption that the future will be like the past is rather dangerous.
However, as we will see below, the recent behaviour of SST
does look rather similar to that of the past. For the whole of the historical SST record, SST is apparently
affected by a trend, by long term variations, and by interannual noise,
and there is no strong indication in the SST data that this has
changed recently. Of course if it has, then our method will be misled. In that case,
one has to resort to making assumptions about \emph{how} the processes affecting SST have changed,
which is rather difficult.

We note that we do not consider potential predictability of Atlantic SST due to the effects of ENSO.
It is well understood that ENSO affects SST in the tropical Atlantic,
and one could imagine that by including information about
the current or predicted state of ENSO one might be able to improve the predictions that we describe here.
We, however, are currently focussing on the much longer timescale processes of AMO and trend.
We plan to include ENSO-related predictors at a later date.

Given our interest in hurricane numbers, we start by considering the year-ahead predictability of SST averaged over
the hurricane season, from June to November. We then also briefly investigate shorter periods, for comparison,
and to assess whether results derived for long periods are likely to apply to shorter periods.

\section{Data}

\begin{figure}[!hb]
  \begin{center}
    \scalebox{0.8}{\includegraphics{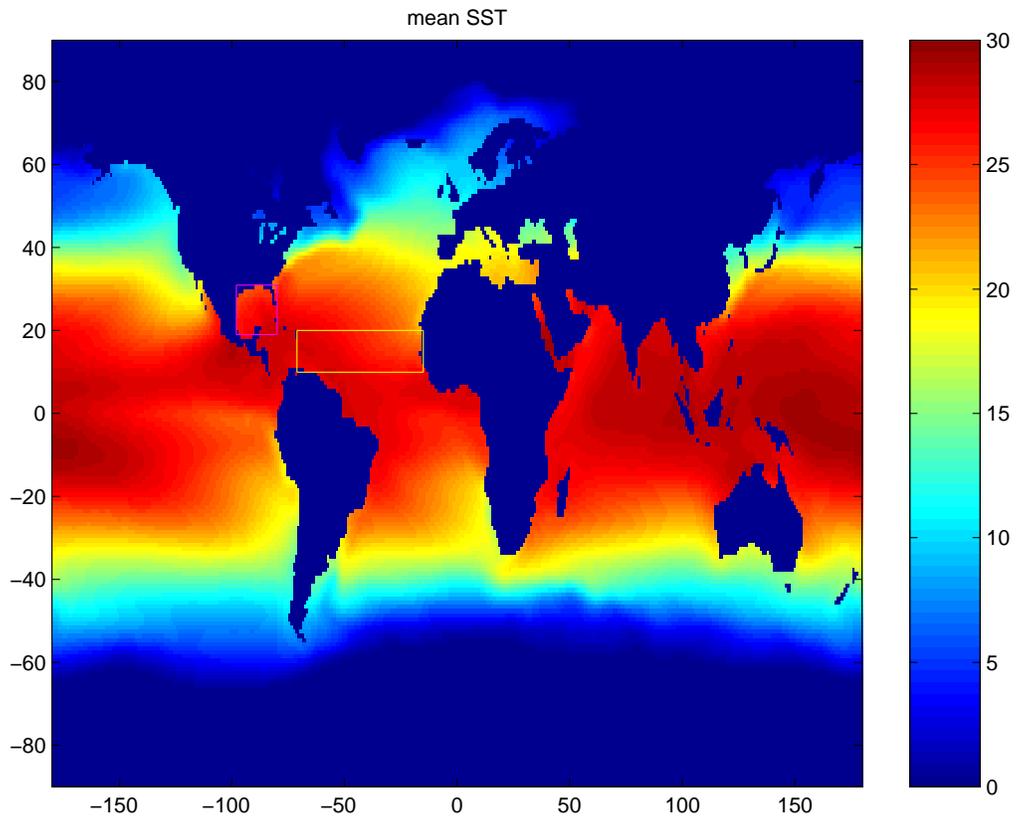}}
  \end{center}
    \caption{
Global mean SST from HadISST.
     }
     \label{f01}
\end{figure}

The data we use for our SST prediction study is a gridded data set known as HadISST~\citep{hadisst}.
This data gives estimates of monthly mean SSTs from 1870 to 2005.
In figure~\ref{f01} we show global SSTs averaged over the whole HadISST data set.
The most striking
pattern in this figure is that the SST in the tropics is much warmer than the SST in the
extratropics and near the poles, as one might expect. There are also other large-scale variations in SST,
such as colder SSTs off the west coasts of major continents.
We are interested in SSTs in the tropical Atlantic, and to that end have defined two regions of
particular interest, shown by boxes in the figure.
The first box is
the Gulf of Mexico, and the second is
the so-called Main Development Region for hurricanes (the MDR), which lies between West Africa and the Caribbean.
We now investigate SST variability in these two regions.

\subsection{Gulf SST}

\begin{figure}
  \begin{center}
    \scalebox{0.6}{\includegraphics{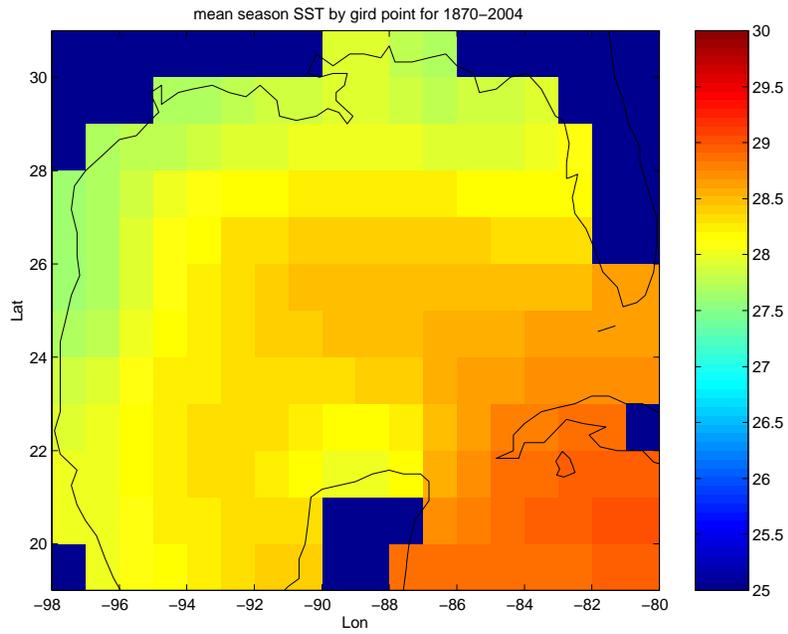}}
  \end{center}
    \caption{
Gulf of Mexico SST for June-November, averaged from 1870 to 2005.
     }
     \label{f02}
\end{figure}

\begin{figure}
  \begin{center}
    \scalebox{0.6}{\includegraphics{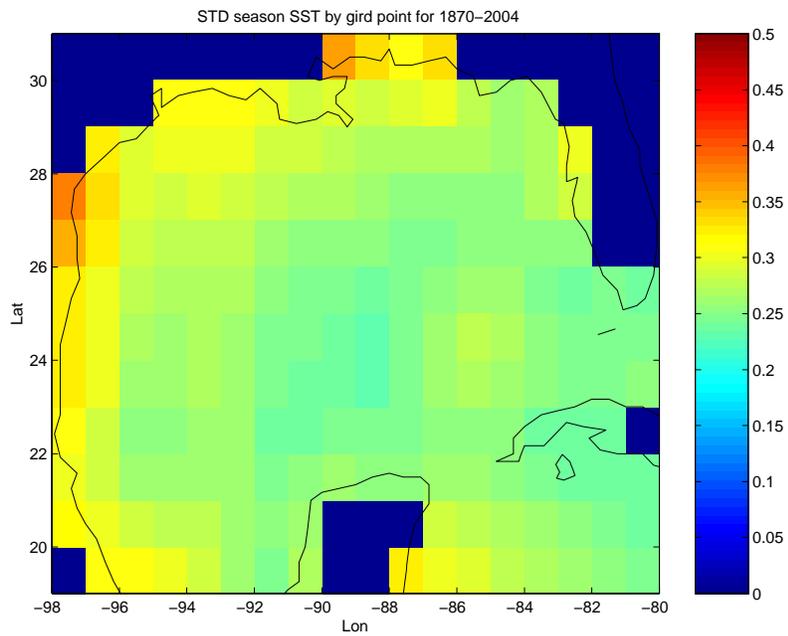}}
  \end{center}
    \caption{
Gulf of Mexico SST for June-November, standard deviation from 1870 to 2005.
}
     \label{f03}
\end{figure}

Figure~\ref{f02} shows the hurricane season mean SST for the period 1870 to 2005 for our Gulf region.
Typical SSTs in this
region are around 28$^{o}$C, with a large-scale gradient from the warmest SSTs in the South East of
the region, to the coolest SSTs in the North West. Figure~\ref{f03} shows the standard deviation
of hurricane season SST in this region from year to year. Typical variations for most of the region are around a quarter
of a degree, with slightly more variability along the North West boundary of the region.

\begin{figure}[!hb]

  \begin{center}
    \scalebox{0.5}{\includegraphics{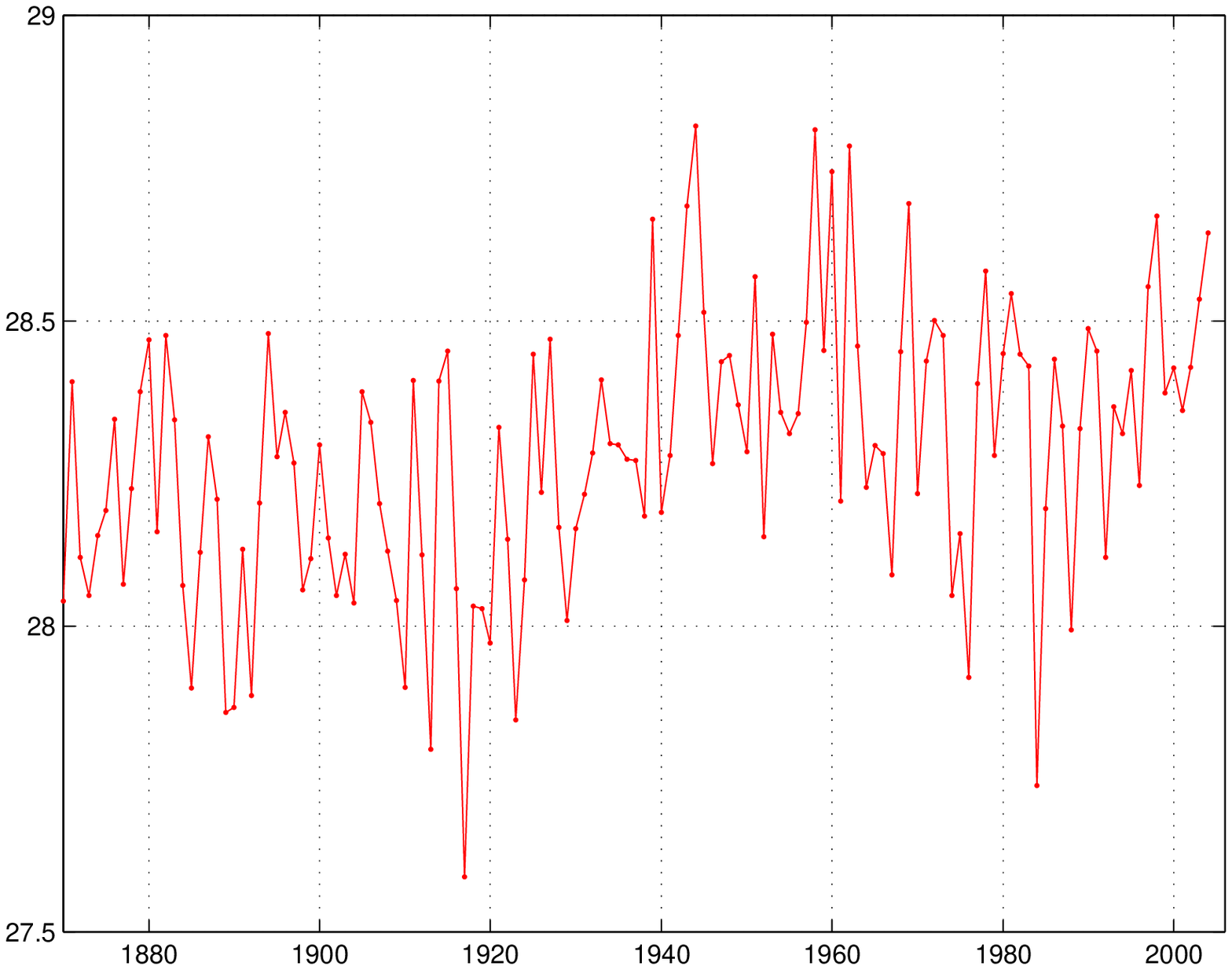}}
  \end{center}
    \caption{
June-November average Gulf of Mexico SST by year from 1870 to 2005.
     }
     \label{f04}

  \begin{center}
    \scalebox{0.5}{\includegraphics{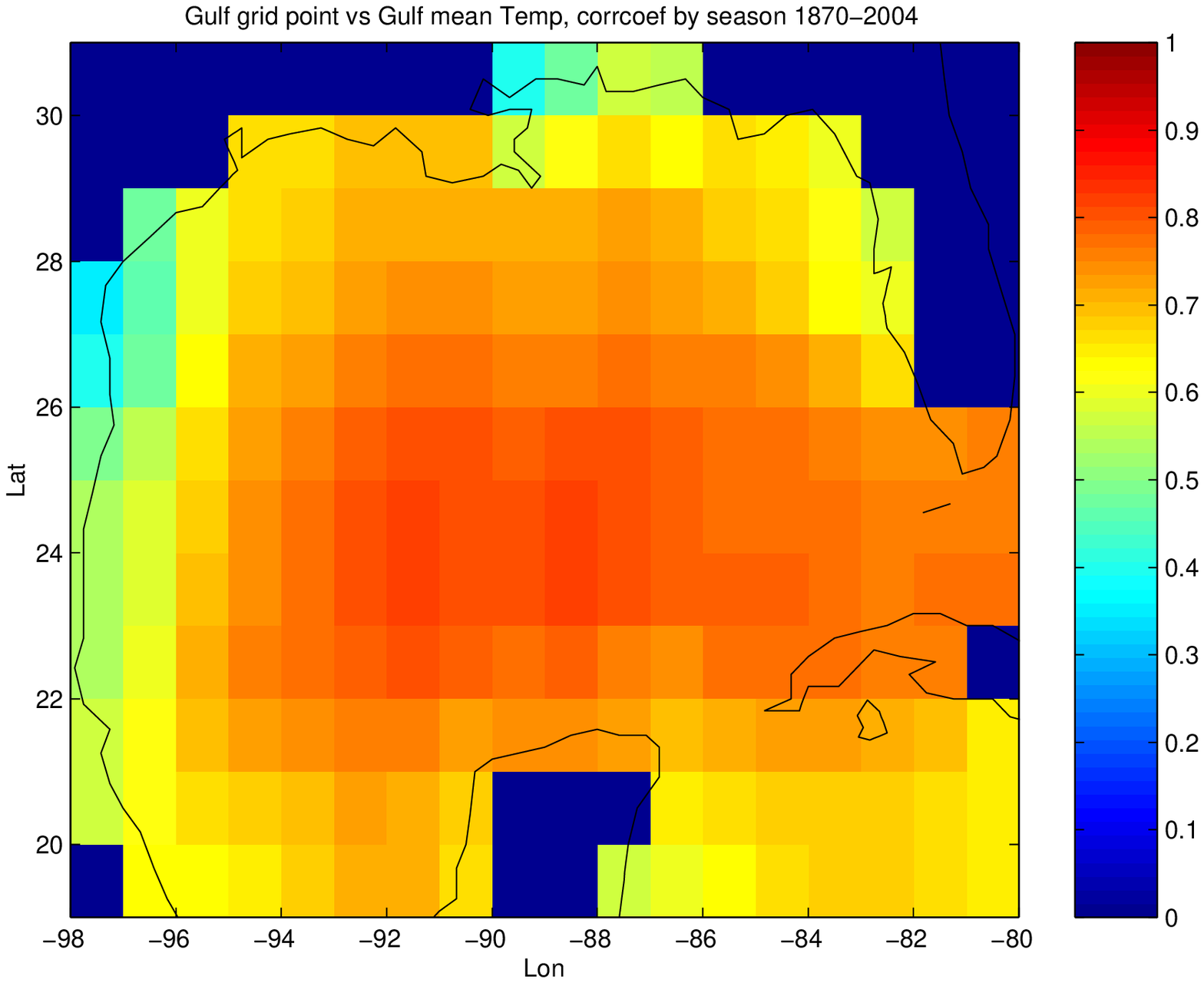}}
  \end{center}
    \caption{
Correlation between the index shown in figure~\ref{f04} and the local June-November SST.
     }
     \label{f05}

  \begin{center}
    \scalebox{0.5}{\includegraphics{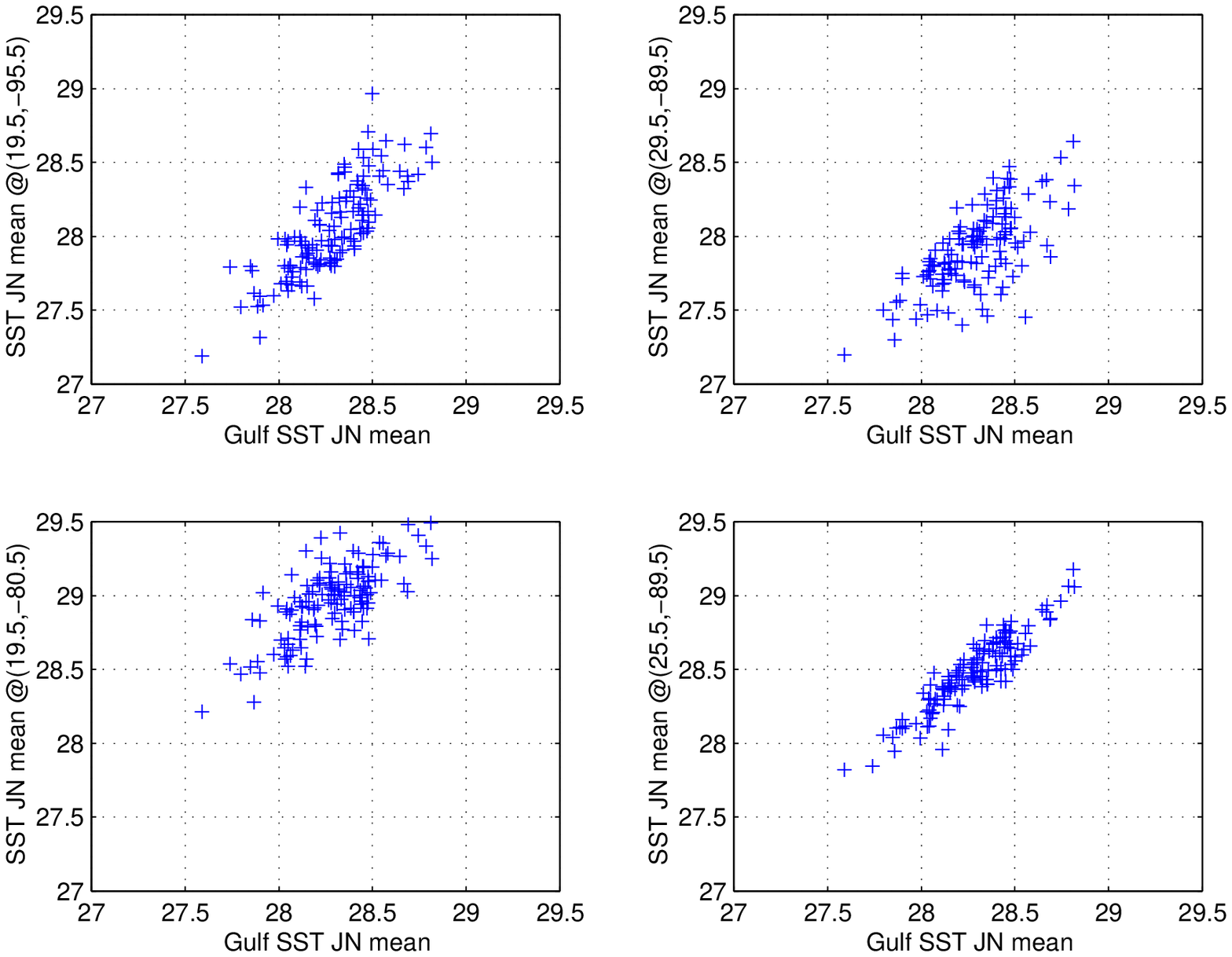}}
  \end{center}
    \caption{
Scatter plots showing the SST index from figure~\ref{f04} (horizontal axis) against
local SST.
     }
     \label{f06}
\end{figure}

In order to render our SST prediction question slightly more tractable,
and as a first step, we average the SSTs in our Gulf region into a single index, with one value per year,
based on the months of the hurricane season. A time series of this index is shown in figure~\ref{f04}.
This time series shows an overall warming trend, multidecadal variability, and interannual variability.
In recent years we see a strong warming trend, but this trend is not unique:
it is very similar to the warming trend present during the 1920s and 1930s.
Does this index really represent the variability within
the whole of the Gulf, or does it throw away much of the detail?
To investigate this question, figure~\ref{f05} shows the linear correlation coefficient between
this index and the local SST values. We see that in the central part of the basin, there is a high correlation
with the index, of around 0.8. In this region, we conclude that SSTs fluctuate in a fashion that is
reasonably well coordinated with our index.
In the boundaries of our Gulf region, however, the correlation
is somewhat lower, dropping to slightly over 0.6. This suggests that our single index is less useful
for describing SST variability in these regions, and that there are significant variations in SST along
the boundaries that are independent of the basin average. We feel that, overall, this correlation
structure is sufficient to justify the use of a single spatially averaged index as a first step, but
does suggest that at a later stage we may need to consider a more detailed analysis.
Figure~\ref{f06} shows these correlations in more detail using scatter plots. In each plot the horizontal
axis shows our SST index, while the vertical axes show SST at 4 points selected from the Gulf region
(the exact locations are given in the labelling on the vertical axis of each panel). Again, we see that
SSTs in the centre of the region are most highly correlated with the index.

\subsection{MDR SST}

\begin{figure}[!t]
  \begin{center}
    \scalebox{0.6}{\includegraphics{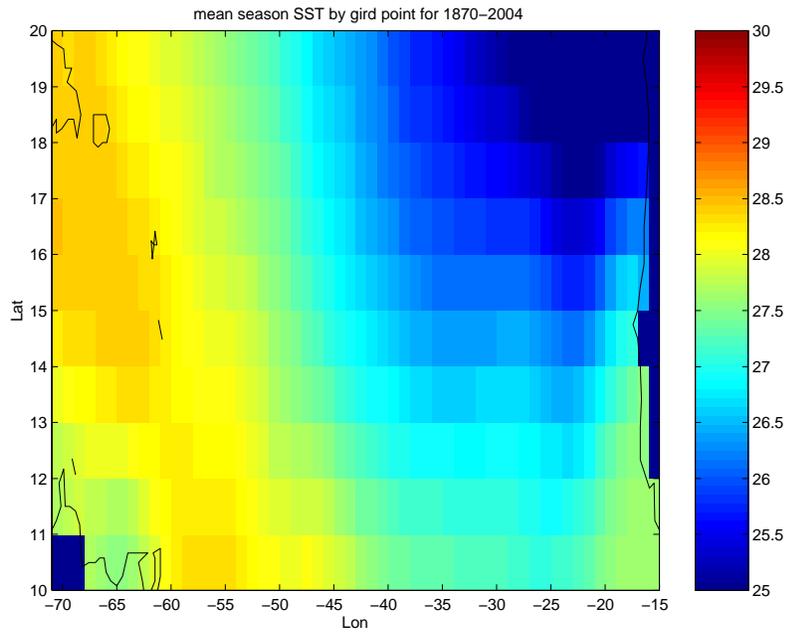}}
  \end{center}
    \caption{
MDR SST for June-November, averaged from 1870 to 2005.
     }
     \label{f07}
\end{figure}

\begin{figure}
  \begin{center}
    \scalebox{0.6}{\includegraphics{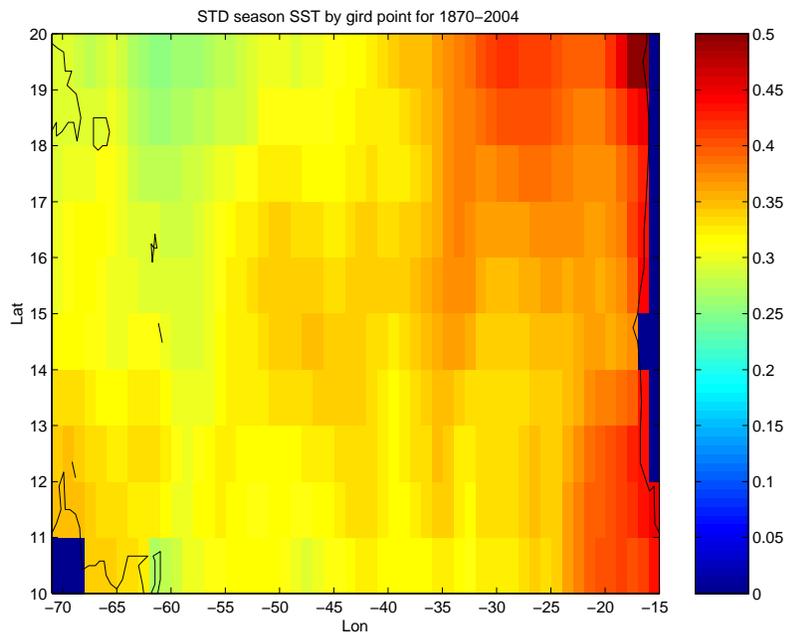}}
  \end{center}
    \caption{
MDR SST for June-November, standard deviation for 1870 to 2005.
     }
     \label{f08}
\end{figure}

\begin{figure}[!hb]
  \begin{center}
    \scalebox{0.5}{\includegraphics{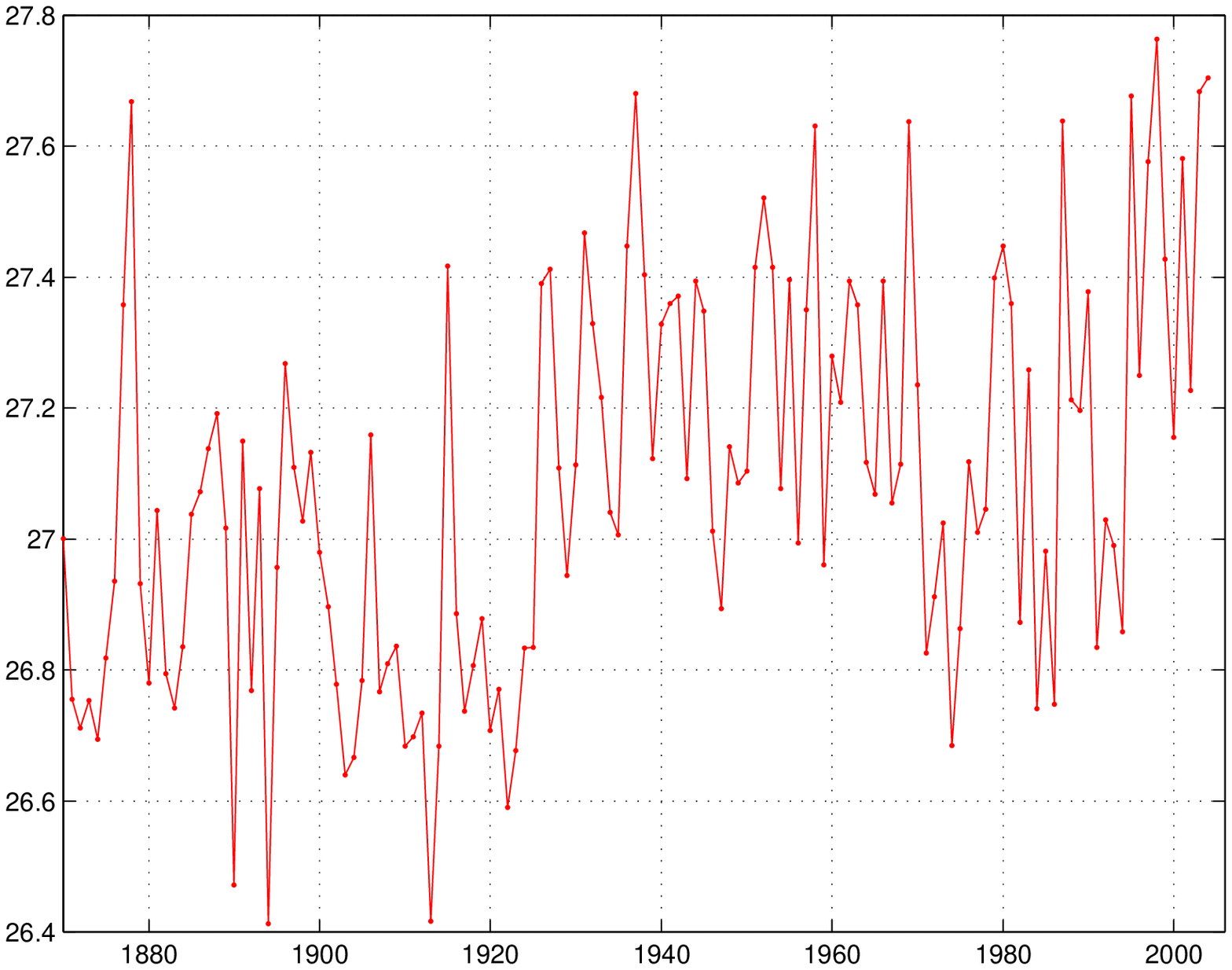}}
  \end{center}
    \caption{
June-November average MDR SST by year from 1870 to 2005.
     }
     \label{f09}

  \begin{center}
    \scalebox{0.5}{\includegraphics{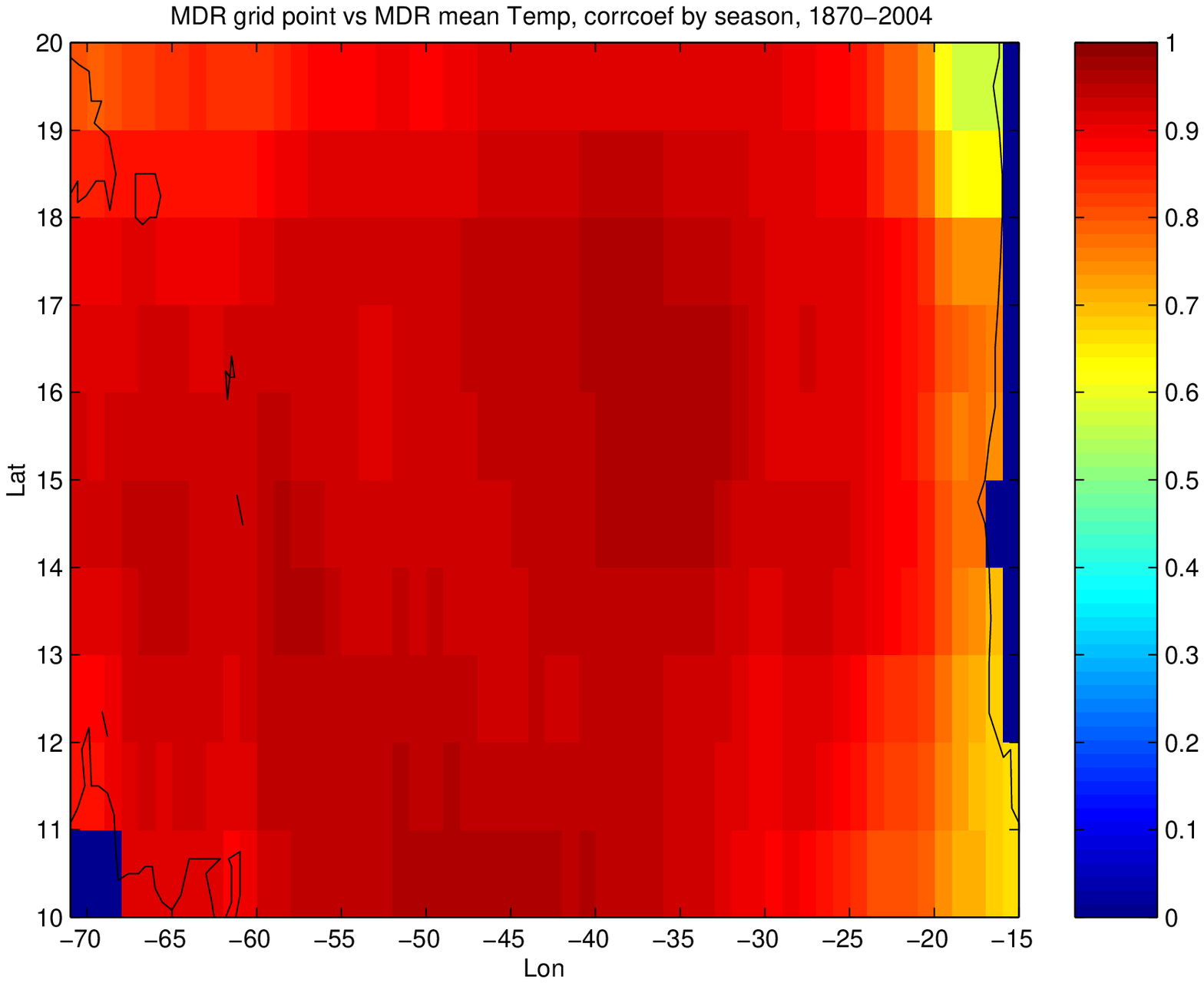}}
  \end{center}
    \caption{
Correlation between the index shown in figure~\ref{f09} and the local June-November SST.
     }
     \label{f10}

  \begin{center}
    \scalebox{0.5}{\includegraphics{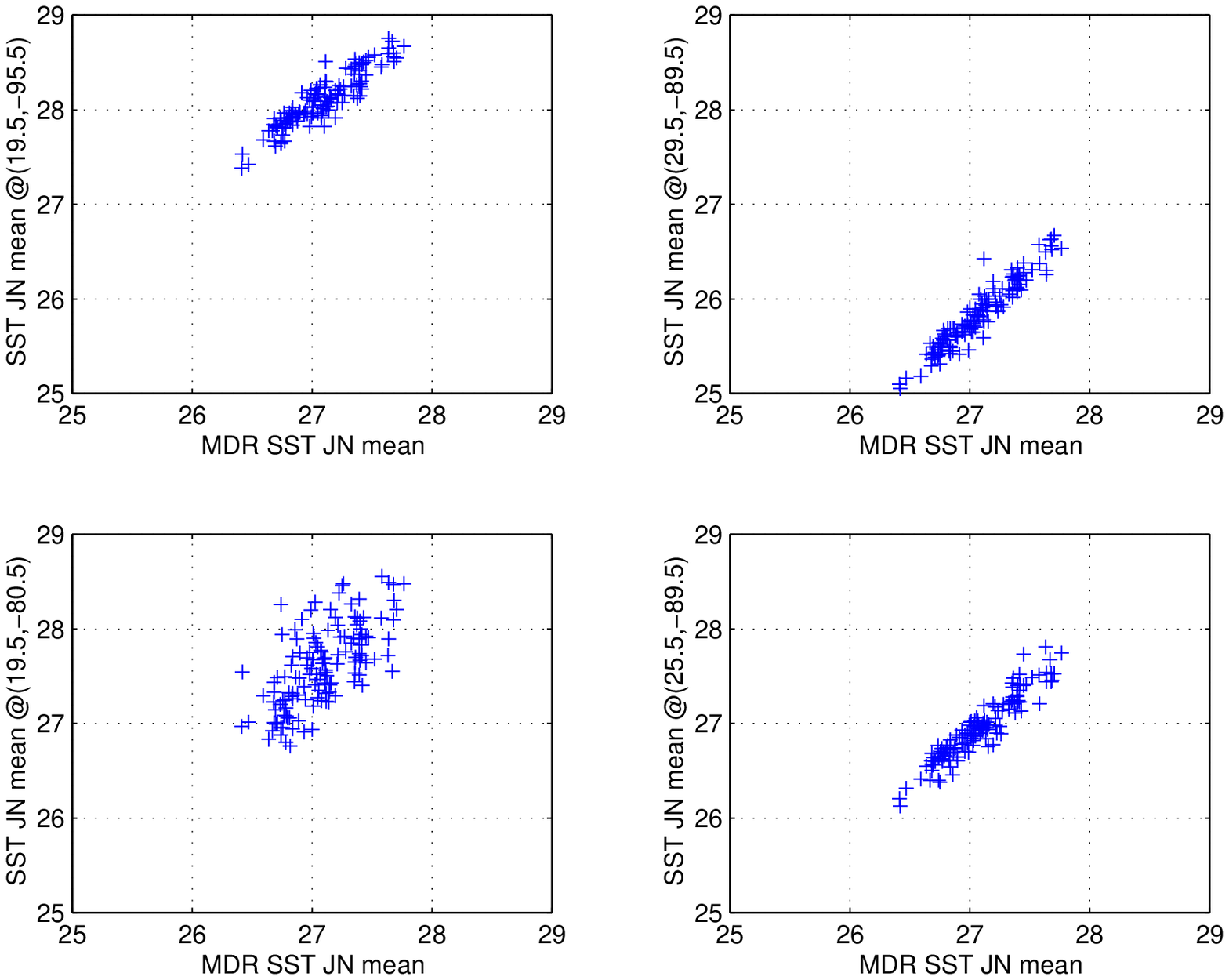}}
  \end{center}
    \caption{
Scatter plots showing the SST index from figure~\ref{f09} (horizontal axis) against
local SST.
     }
     \label{f11}
\end{figure}

Figure~\ref{f07} shows the hurricane season mean SST for our MDR region. There is a considerable
variation in mean SST across this region, from waters as cold as 25$^{o}$C in the North East to
waters warmer than 28$^{o}$C in the West. The year to year variability is illustrated in
figure~\ref{f08}; the highest variability is in the East, while temperatures in the West are
more constant.
Comparison with figure~\ref{f03} shows that the variability in SST in this region is higher than that
in the Gulf.
As for the Gulf, we now define an index as the hurricane season average temperature
over this region. This index is shown in figure~\ref{f09}.
Again, the index shows a long term warming trend, interdecadal and interannual variability. The MDR and
Gulf indices are somewhat similar in terms of the shape of the long-term variability:
for instance, both show warming in the 1920s and 1930s, and cooling in the 1960s.
When we correlate the MDR index with
the local temperatures, we find higher correlations than in the Gulf: almost the entire region
shows correlations above 0.8, and much of it above 0.9. We conclude that the SST in this region
to a great extent moves as one. This is good justification for the use of a single index
to summarise the year to year variability. Figure~\ref{f10} confirms this by showing
correlations between the index and individual points: 3 out of 4 of the points
chosen show very high correlation with the MDR index. Only the point in the extreme North East of
the domain is not highly correlated.

\section{Method}

As discussed in the introduction, our plan is to test a number of simple statisical methods for predicting
SST. We have now reduced the problem to predicting two SST indices that are representative of the
SST variability in the two regions we are considering.
Whichever prediction methods do best, we will use to make SST predictions
for the future. We will compare 3 methods, which we call \emph{flat-line}, \emph{linear trend}, and
\emph{damped linear trend} (the use of these three methods is taken from a similar problem that
arises in the pricing of weather derivatives:
see~\citet{j58}).
Each of these methods uses the $n$ years of data from year $i-n$ to year
$i-1$ to predict year $i$. We will vary $n$ for each of the methods, to find which values of $n$ would have given the
best results.

\noindent We now describe the 3 methods in more detail:

\subsubsection{Flat-line}

What we call the flat-line (FL) method is the obvious use of a trailing moving average to predict the next year.
As a statistical prediction scheme it has the advantage that there is only a single parameter that needs to
be estimated, and so the effects of estimation error on the accuracy of the final prediction are likely
to be relatively small. The FL method can capture trends and cycles by using a small value of $n$.
However, a small value of $n$ increases the estimation error.

\subsubsection{Linear trend}

Perhaps the most obvious extension of the flat-line method is to a best-fit linear trend (LT), extrapolated one year forward
to give a prediction. Compared to FL, LT has the disadvantage that there are now two parameters
that must be estimated,
and this will add extra error in the final prediction because of the extra estimation uncertainty.
One can say that linear-trends are always more over-fitted than flat-lines. On the other hand, with data
like the SST data we are looking at, one might hope that use of a LT model might capture
the gradual increasing trend in SST, and so might work well.

\subsubsection{Damped linear trend (DLT)}

As discussed above, the linear trend is more over-fitted than flat-line.
In fact, the best-fit LT model is not even an optimal predictor if the real trend is linear, because
of this overfitting: it is only a `best-fit' in an in-sample sense, not in an out-of-sample or predictive sense.
This raises the question of whether one should \emph{ever} use best-fit linear trends for prediction, and whether
there is something in-between FL and LT that might have some of the
benefits of the linear trend model, but
is less overfitted. The answer to this question is to use something that we will call
a `damped linear trend' (DLT), which we take from~\citet{j58}. The DLT model
is the optimal combination of the FL and LT models, and
is the best way to predict a real linear trend (in terms of minimising the root mean square error of the prediction).
DLT can also be interpreted and explained in a number of other ways~\citep{j74}.

The one potential shortcoming of the DLT model is that the damping parameter, that
determines the proportions of flat-line and linear trend that the method uses, has to be estimated.
Given a perfect estimate, damped linear trends are always better than both flat-line and
linear trends. Given an imperfect estimate, they may not be.

\section{Results}

We now show some results from our 3 prediction schemes.
In each case we show the root-mean-square error (RMSE), the bias and the error standard deviation (SD).
Our goal is to make predictions with low RMSE. The bias and error SD, which are the two components
of the RMSE, can help us understand what is driving the RMSE scores from the different models.

\subsection{Gulf SST June-November}

\begin{figure}[!t]

  \begin{center}
    \scalebox{0.6}{\includegraphics{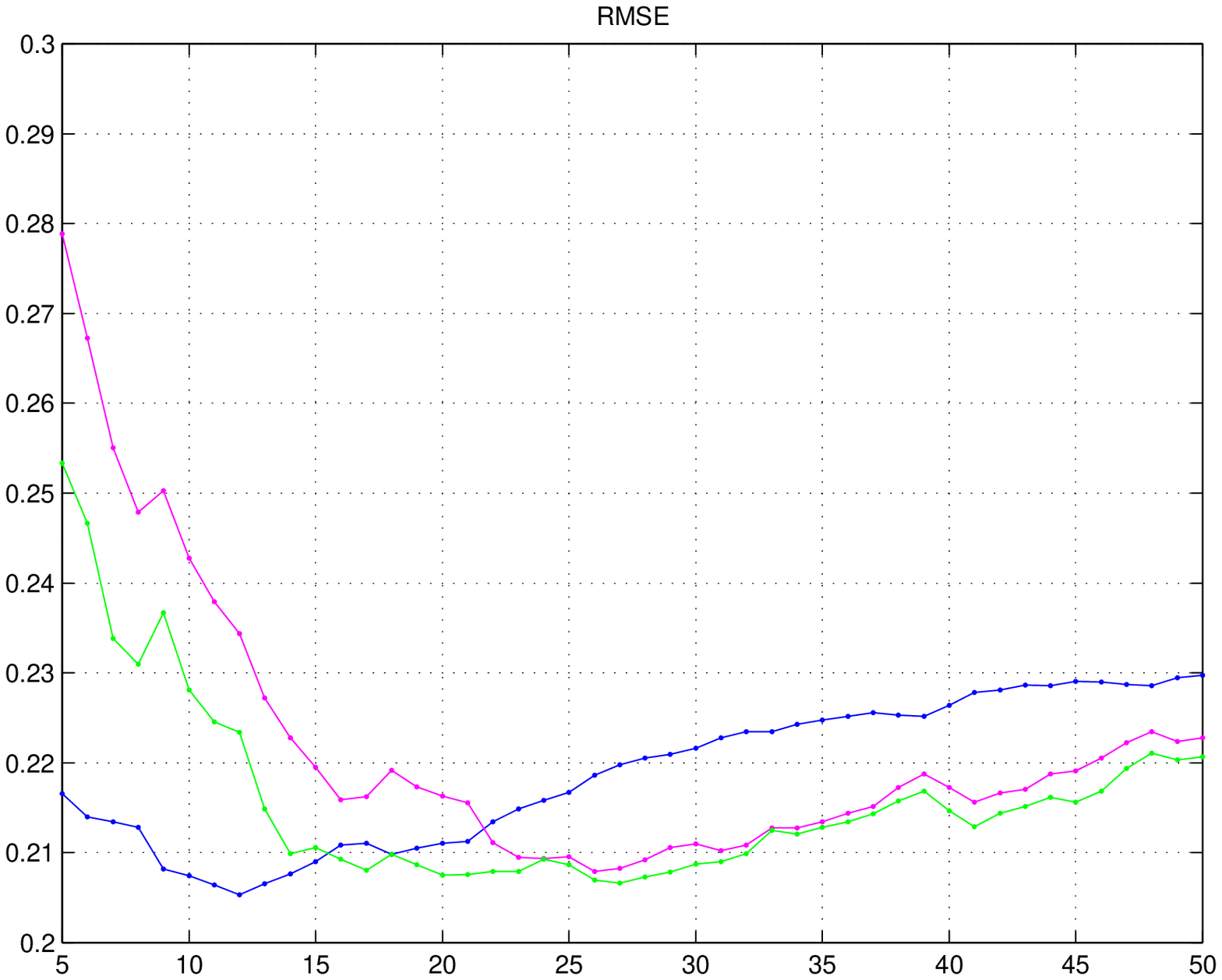}}
  \end{center}
    \caption{
RMSE for year-ahead predictions of the Gulf of Mexico June-November SST index shown in figure~\ref{f04} for
three simple statistical prediction models: flat-line (blue), best fit linear trend (red) and
damped linear trend (green).
     }
     \label{f12}

  \begin{center}
    \scalebox{0.4}{\includegraphics{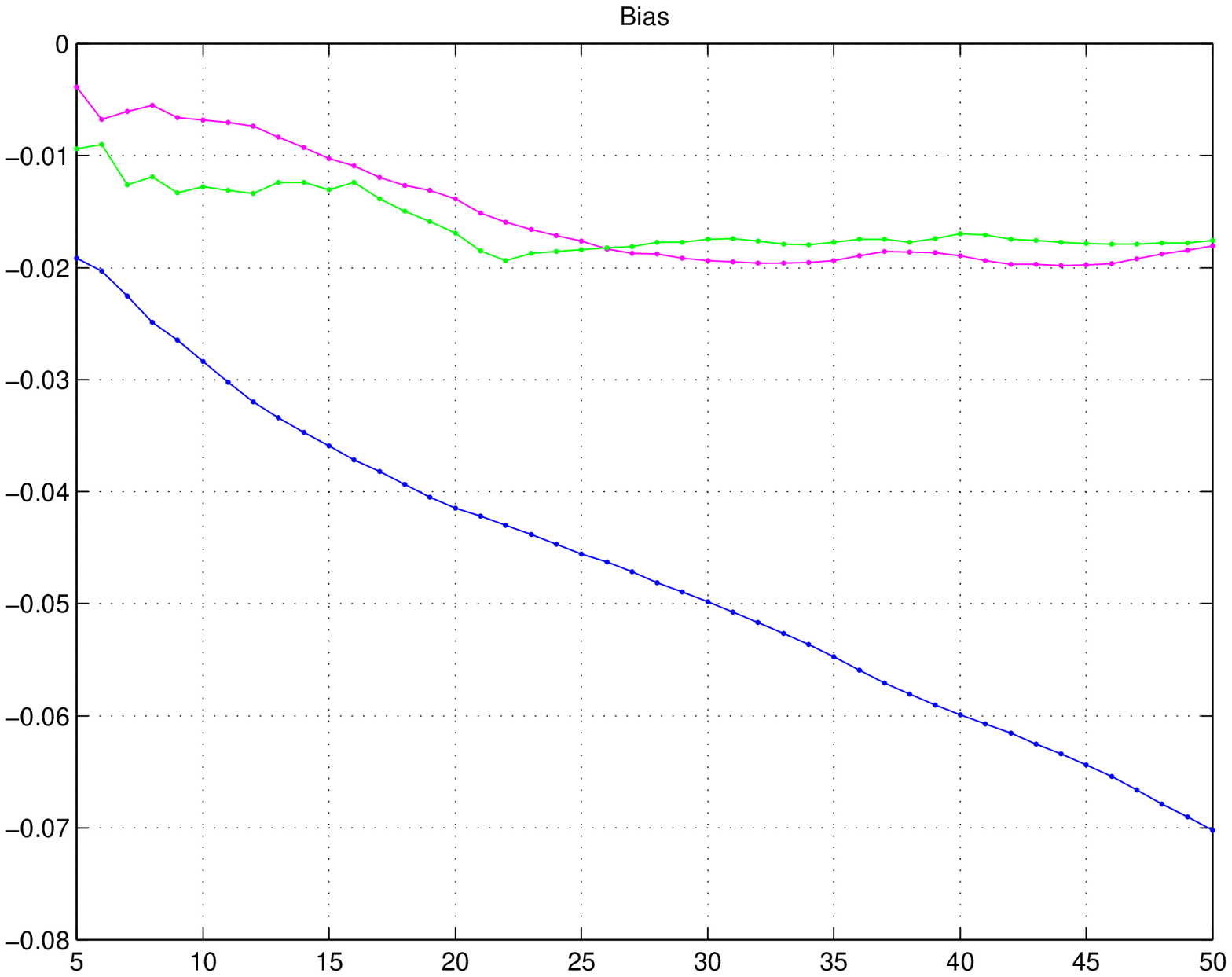}}
  \end{center}
    \caption{
Bias for the three predictions described above.
     }
     \label{f13}

  \begin{center}
    \scalebox{0.4}{\includegraphics{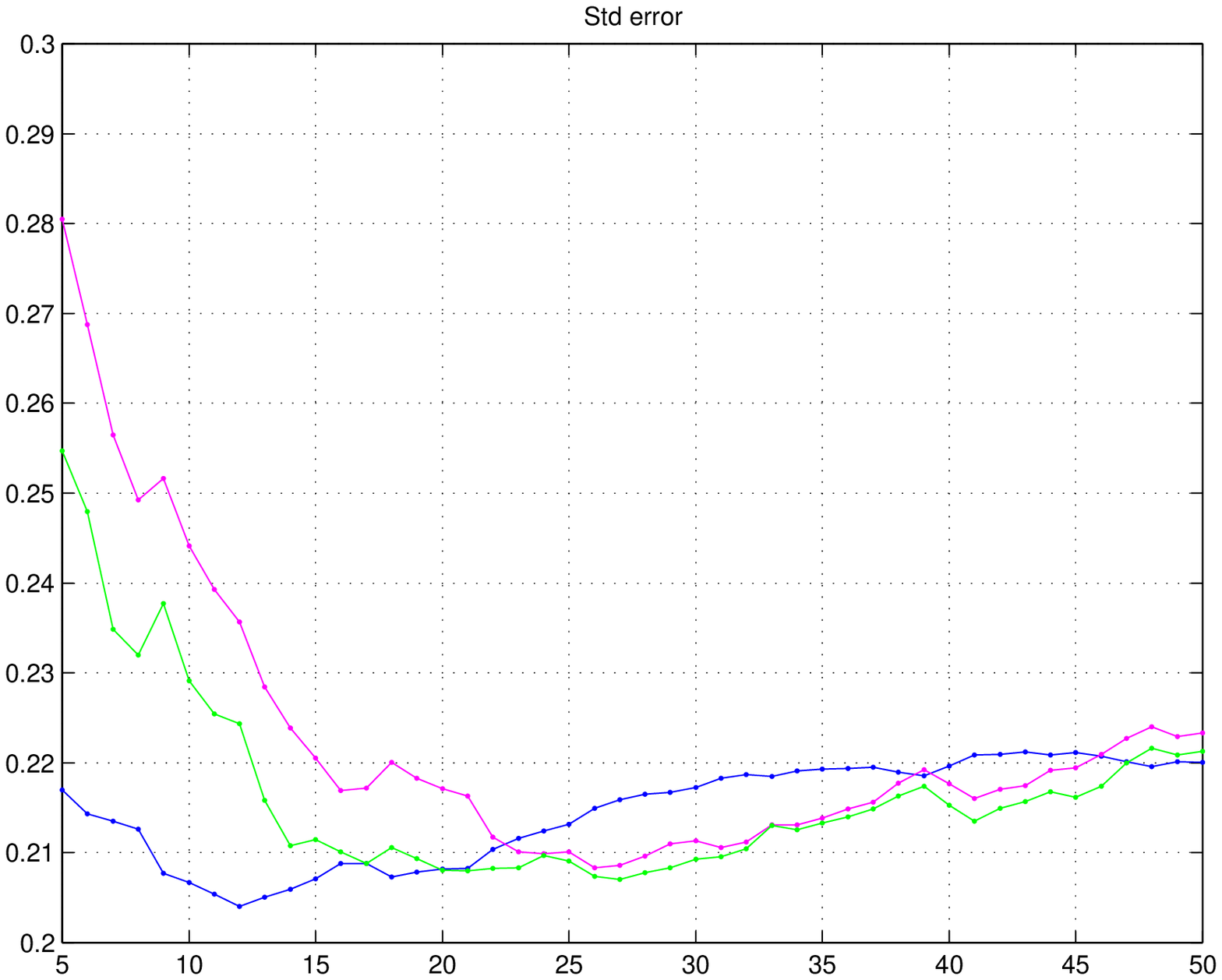}}
  \end{center}
    \caption{
SD of errors for the three predictions described above.
     }
     \label{f14}
\end{figure}

Figure~\ref{f12} shows the MSE scores from the 3 models, versus numbers of years of data used
in each, when used to predict our June-November Gulf SST index.
The blue line shows results from the FL model. We see that the most effective hindcasts from the
FL model are for a window length of 12 years. As the window length increases beyond 12 years the
FL forecasts become progressively worse, with the RMSE increasing more or less linearly. This is presumably because
the FL model ignores the upward trend in the SST index. The pink line shows the results for the LT
model. This model gives very poor forecasts for 15 year windows or less, presumably because the two parameters
of the model are very poorly estimated when such a small amount of data is used. The LT model performs
best at 26 years, but doesn't do quite as well as the FL model at 12 years. If using less than 22 years
of data, the FL model beats LT, while if using more than 21 years of data LT beats
FL. The green line shows the results for the DLT model. This model performs best with
27 years of data, and achieves a minimum which is slightly lower than that achieved by the
LT model, but slightly higher than that achieved by the FL model.
The bias and SD results for the 3 models show that the RMSE is dominated by the SD, except in the FL model
for large $n$, where the bias is also large enough to contribute materially to the RMSE.

If using more than 16 years of data the DLT is the best of the three models. For fewer years of
data it is beaten by the FL model.
Which, then, is the best of the models? It seems that the \emph{worst} model of the three is the LT.
We say this because the LT model performs the worst of the three numerically and
only competes at all with the best of the other two methods for larger numbers of years of data.
In general, methods that do well for fewer years of data are more useful because they are more likely
to be able to adapt to recent signals, such as an increased rate of climate change, that were not present in the
earlier data.
Choosing between FL and DLT is harder. The best performance of the two is not
materially different. DLT is perhaps slightly better, because the minimum in RMSE is broader,
and hence DLT will give good results over a range of window lengths (and so is less sensitive to
the wrong choice of window length).

\begin{figure}[!hb]
  \begin{center}
    \scalebox{0.8}{\includegraphics{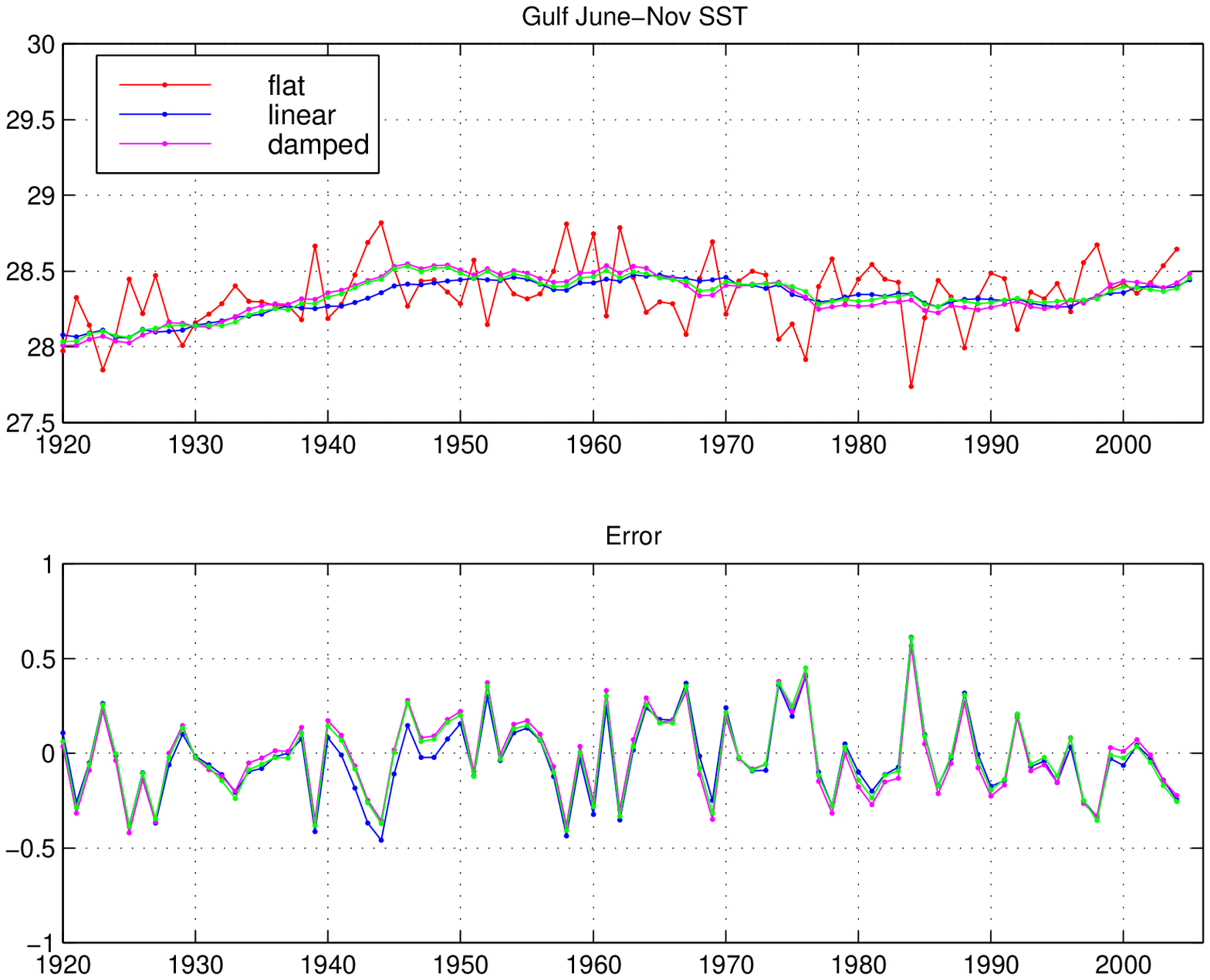}}
  \end{center}
    \caption{
The top panel shows hindcasts for the SST index shown in figure~\ref{f04}
from the flat-line (blue), best fit linear trend (red) and damped linear trend (green) models,
along with actual values for the index.
The lower panel shows the errors from each of the three predictions.
     }
     \label{f15}
\end{figure}

The upper panel of figure~\ref{f15} shows hindcasts made from the 3 methods using the best window lengths from each.
We see that the three predictions are not vastly different, and the FL and DLT predictions
are the closest pair. The errors from the 3 different methods (shown in the lower panel of figure~\ref{f15})
are very similar, and are driven by
interannual variability, including the effects of ENSO, that is not captured by any of the schemes.
The errors do not show any significant trend, decadal or multidecadal variability.


\subsection{MDR SST June-November}
\begin{figure}[!hb]

  \begin{center}
    \scalebox{0.6}{\includegraphics{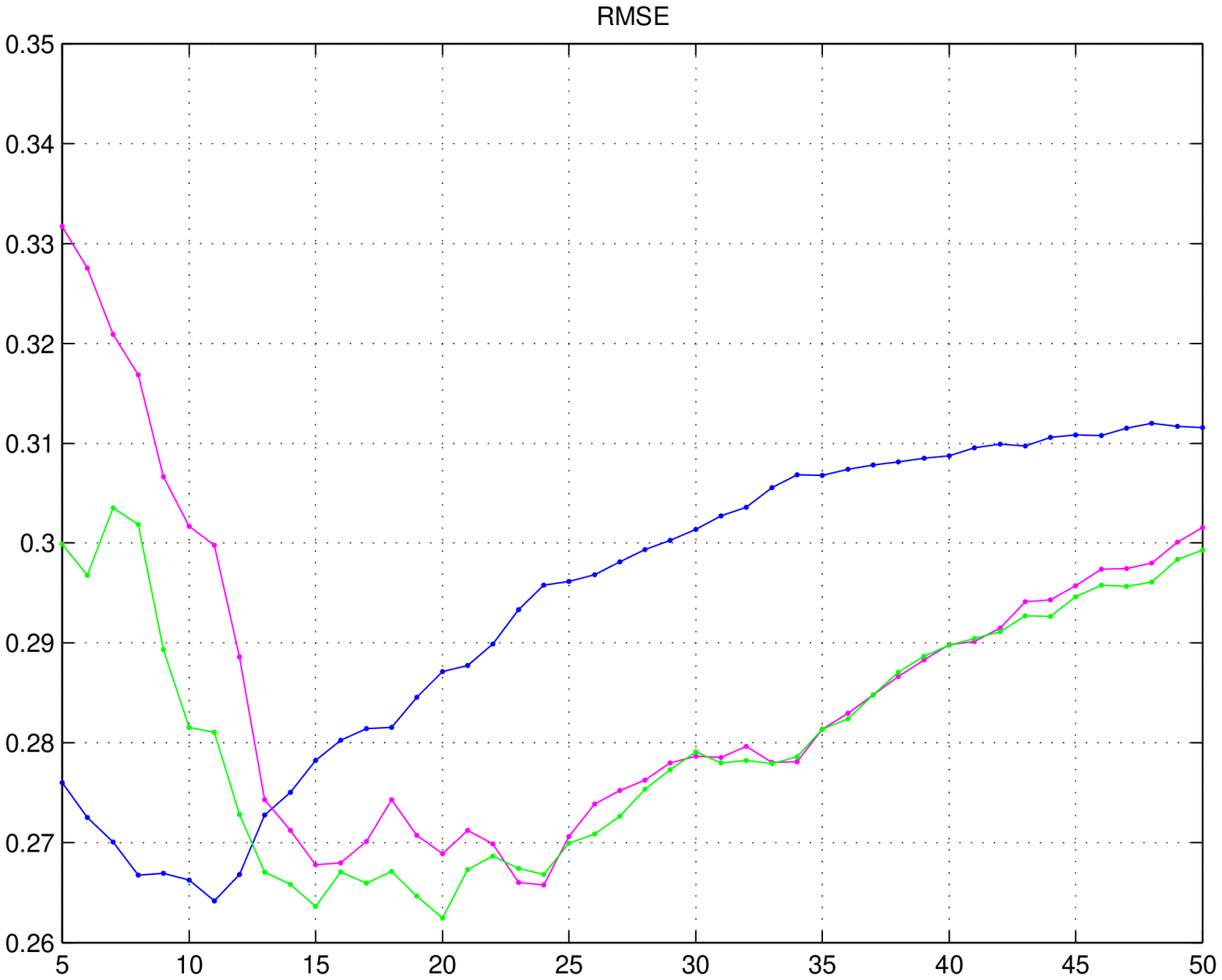}}
  \end{center}
    \caption{
RMSE for year-ahead predictions of the MDR June-November SST index shown in figure~\ref{f09} for
three simple statistical prediction models: flat-line (blue), best fit linear trend (red) and
damped linear trend (green).
     }
     \label{f16}

  \begin{center}
    \scalebox{0.4}{\includegraphics{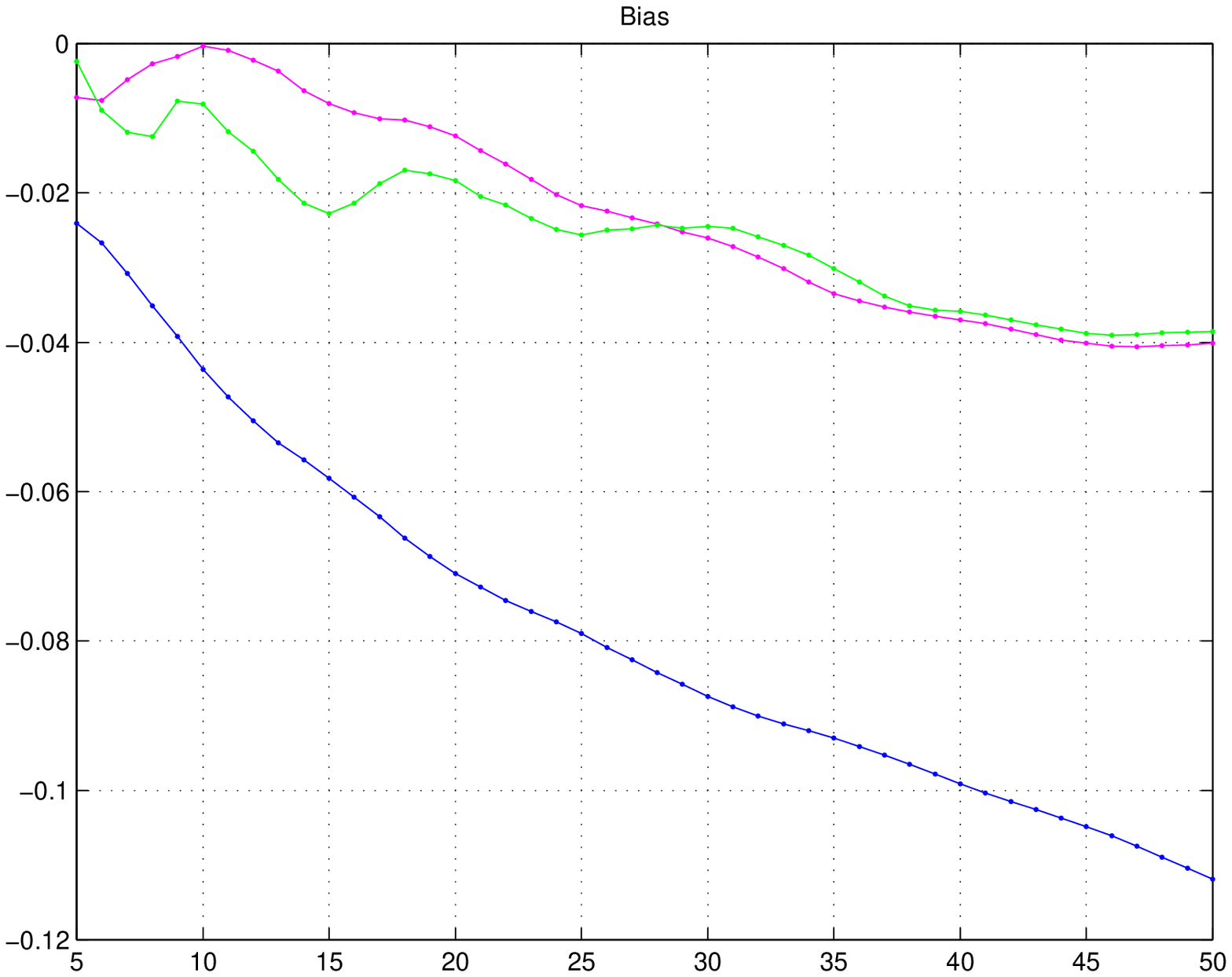}}
  \end{center}
    \caption{
Bias for the three predictions described above.
     }
     \label{f17}

  \begin{center}
    \scalebox{0.4}{\includegraphics{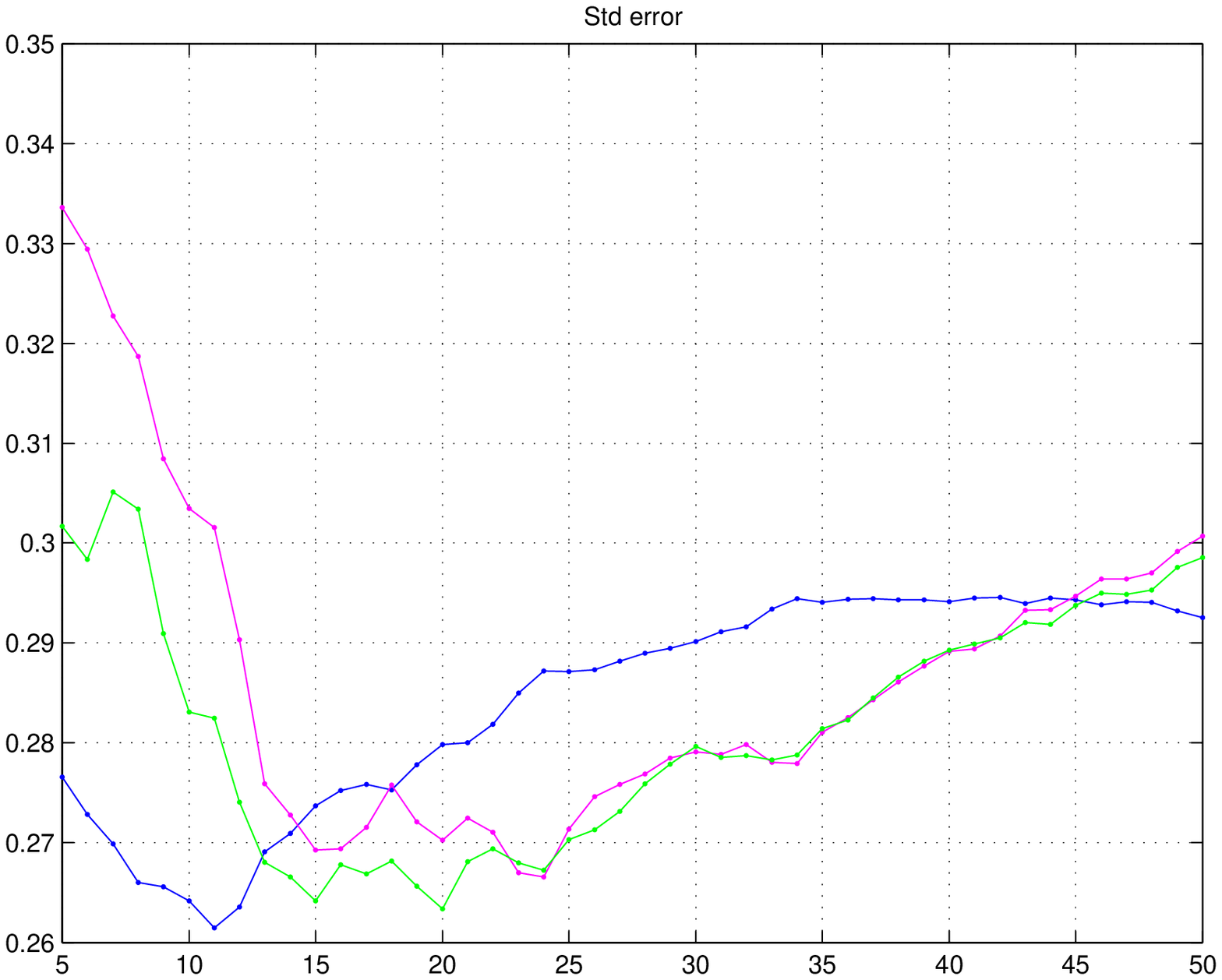}}
  \end{center}
    \caption{
SD of errors for the three predictions described above.
     }
     \label{f18}

\end{figure}

Figure~\ref{f16} shows the results for predictions of MDR SSTs.
The first thing we notice is that the predictions for MDR SST are worse than those for Gulf SST.
For the Gulf SSTs, the best predictions all had MSE values less than 0.21$^{o}$C. For the MDR
region the best predictions have errors that are roughly 50\% higher: this is presumably
related to the higher level of interannual variability in this region noted earlier.
The FL model does its best with a short window of 11 years, and deteriorates rapidly for
longer windows.
The LT model does its best for a 24 year window, and does roughly as well
as the FL model. The DLT model gives the best results of the three,
with the optimal predictions coming from window lengths of 20 years, and again we see that
the sensitivity to window width around the point of minimum RMSE is the lowest.
For less than 13 years of data the FL model does best. For 13-22 years of data
the DLT model does best, while for longer than 22 years of data
DLT and LT models are roughly the same.


\subsection{Gulf SST August-September}

\begin{figure}[!hb]
  \begin{center}
    \scalebox{0.6}{\includegraphics{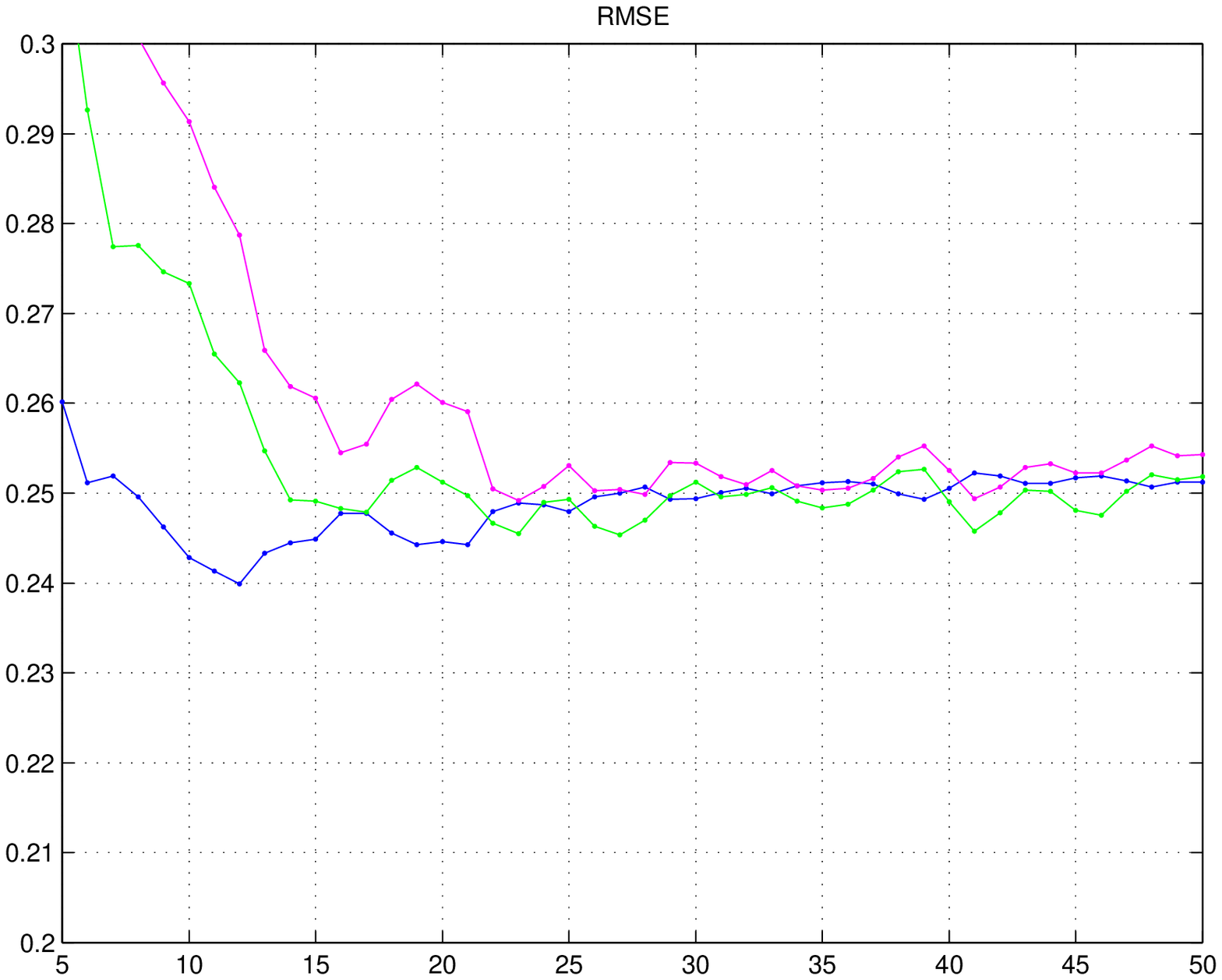}}
  \end{center}
    \caption{
RMSE for year-ahead predictions of a Gulf of Mexico August-September SST index using
three simple statistical prediction models: flat-line (blue), best fit linear trend (red) and
damped linear trend (green).
     }
     \label{f20}

  \begin{center}
    \scalebox{0.4}{\includegraphics{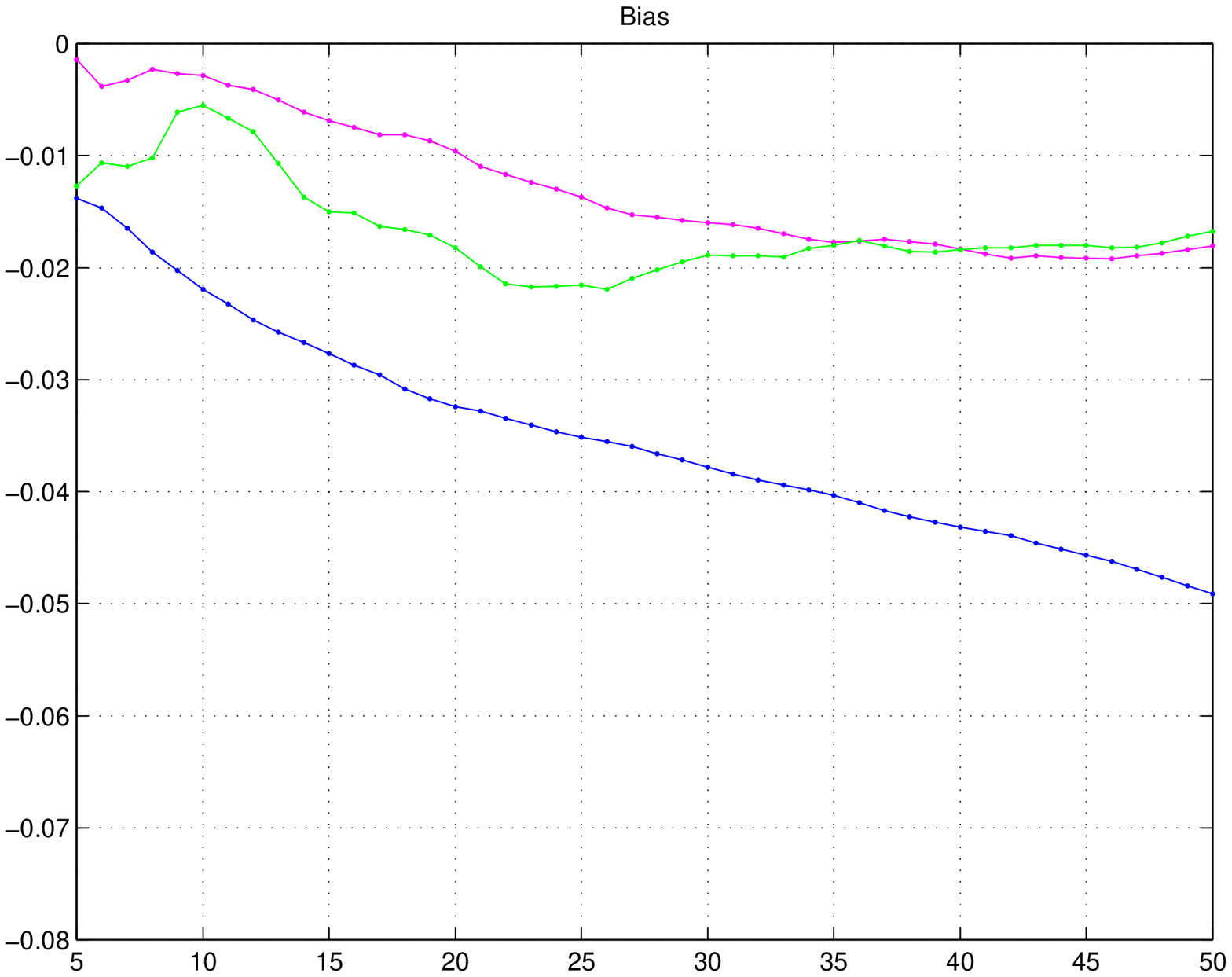}}
  \end{center}
    \caption{
Bias for the three predictions described above.
     }
     \label{f21}

  \begin{center}
    \scalebox{0.4}{\includegraphics{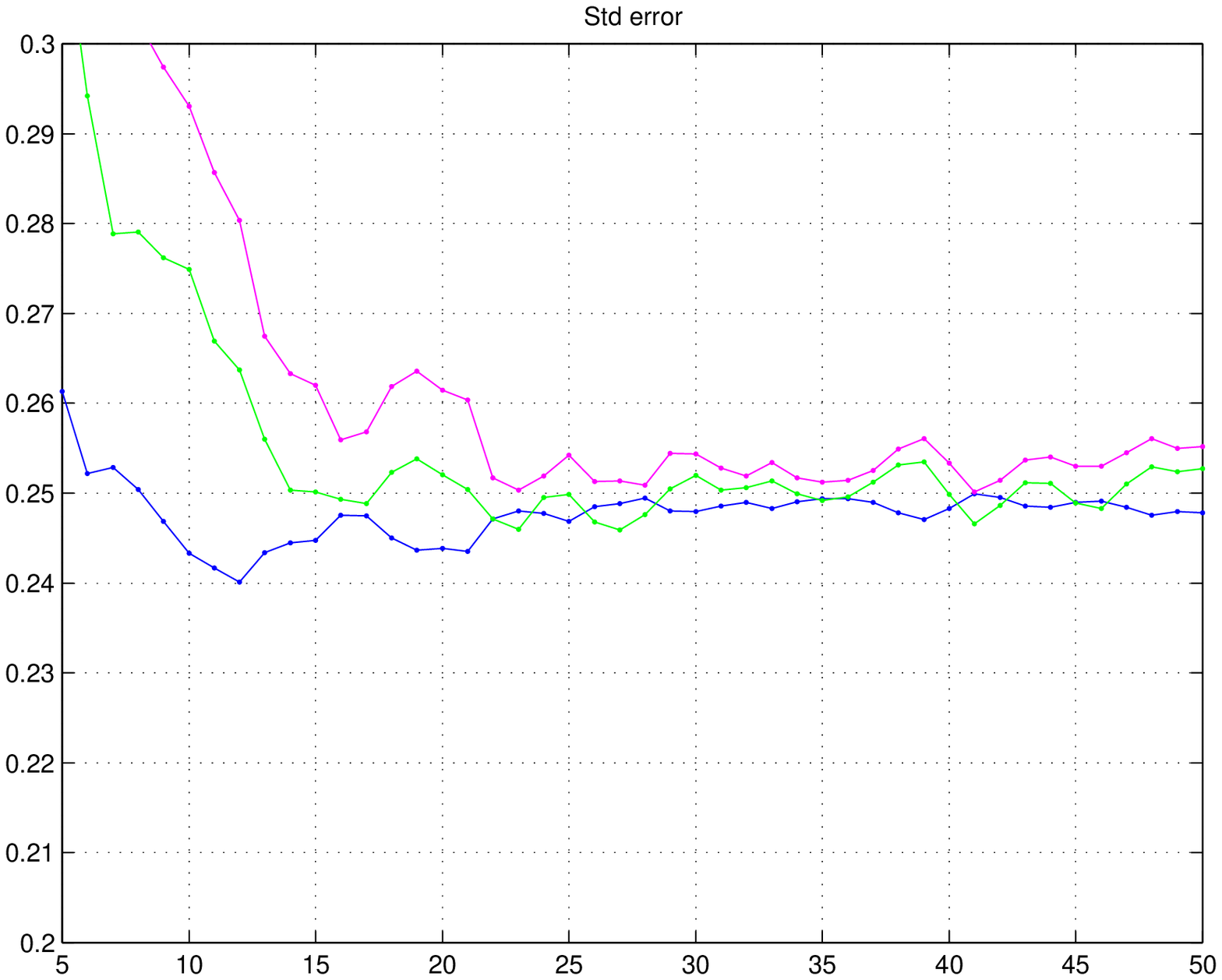}}
  \end{center}
    \caption{
SD of errors for the three predictions described above.
     }
     \label{f22}

\end{figure}

\begin{figure}[!hb]
  \begin{center}
    \scalebox{0.8}{\includegraphics{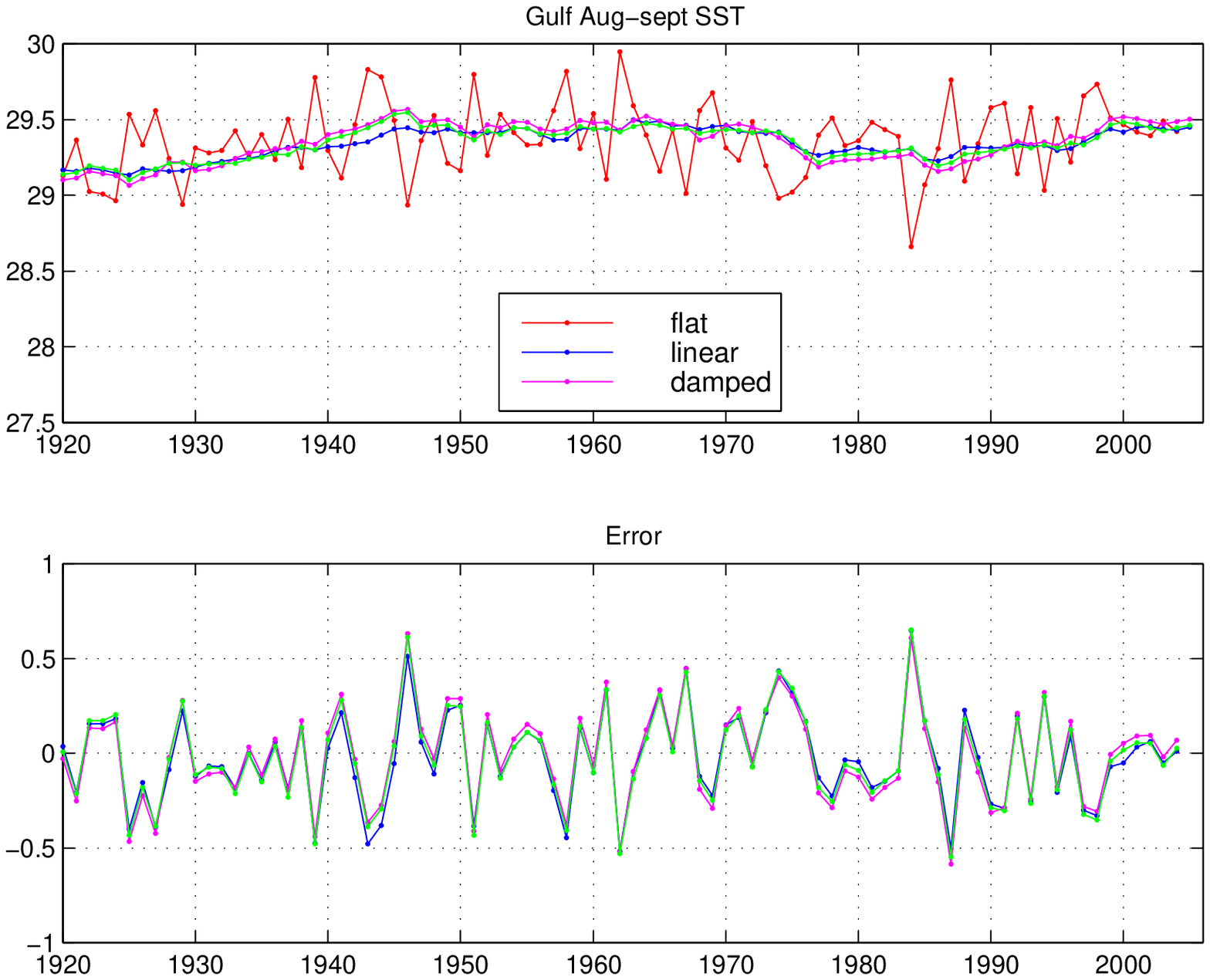}}
  \end{center}
    \caption{
The top panel shows hindcasts for the August-September Gulf SST index.
from the flat-line (blue), best fit linear trend (red) and damped linear trend (green) models,
along with actual values for the index.
The lower panel shows the errors from each of the three predictions.
     }
     \label{f23}
\end{figure}

Although our principle focus is on the June-November period, we now briefly show results
for August-September (in figures~\ref{f20} to~\ref{f23})
to get some idea whether the results for June-November are likely to hold
for shorter time periods. In fact, the results are rather different. The FL model
performs the best, the DLT is close behind, and the LT model performs badly.
The errors are larger than those for June-November.

\subsection{MDR SST August-September}

\begin{figure}[!hb]

  \begin{center}
    \scalebox{0.6}{\includegraphics{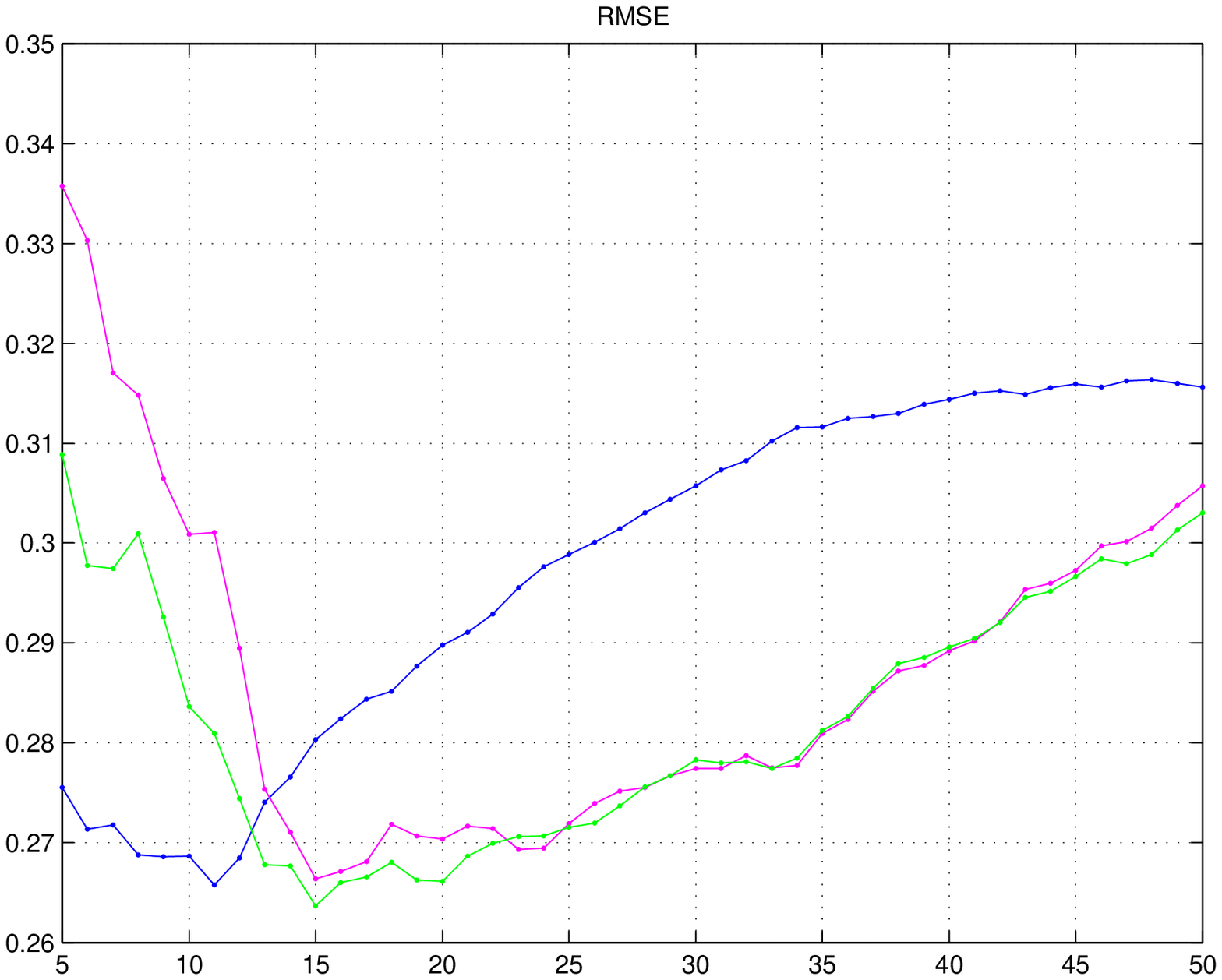}}
  \end{center}
    \caption{
RMSE for year-ahead predictions of an MDR August-September SST index for
three simple statistical prediction models: flat-line (blue), best fit linear trend (red) and
damped linear trend (green).
     }
     \label{f24}

  \begin{center}
    \scalebox{0.4}{\includegraphics{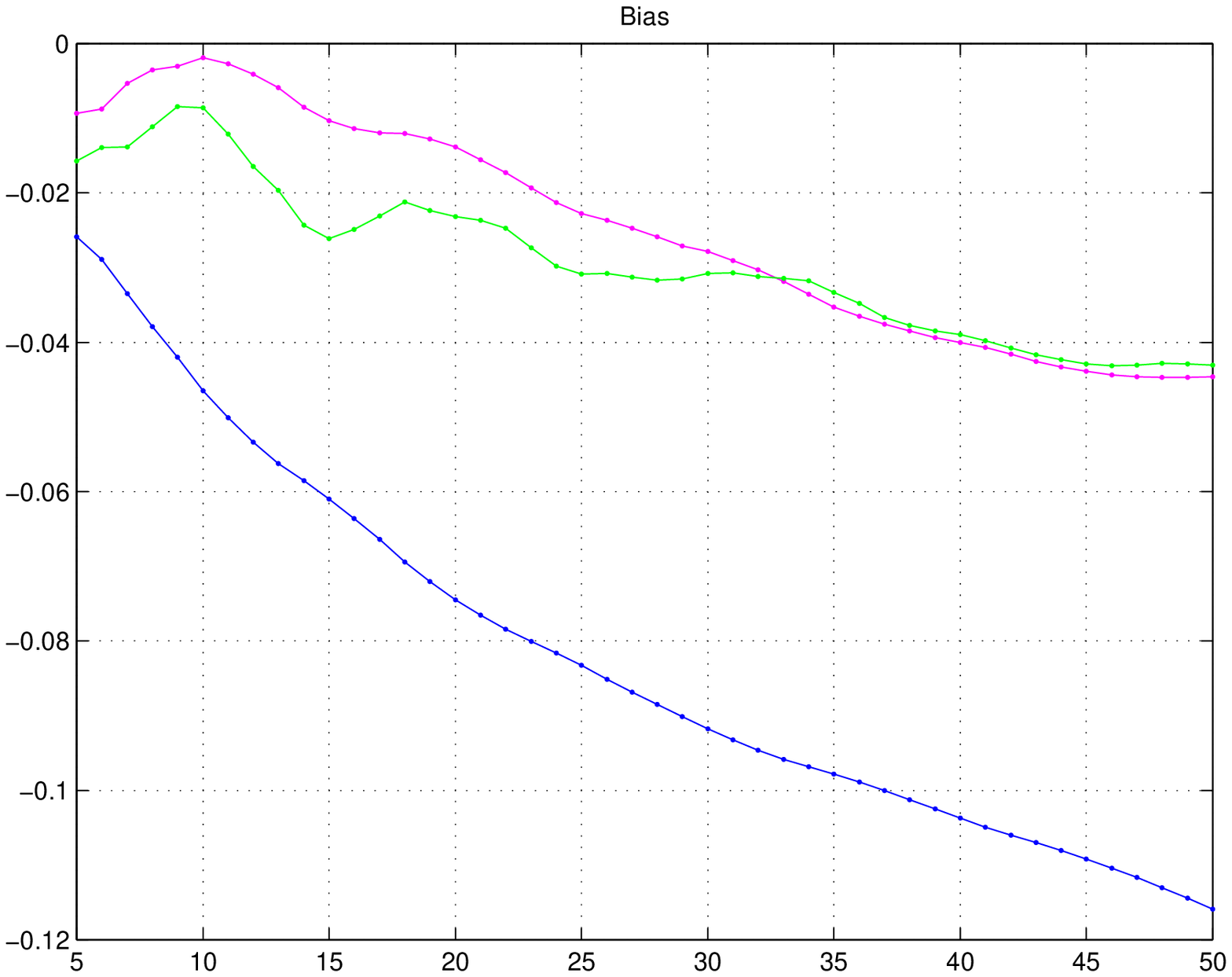}}
  \end{center}
    \caption{
Bias for the three predictions described above.
     }
     \label{f25}

  \begin{center}
    \scalebox{0.4}{\includegraphics{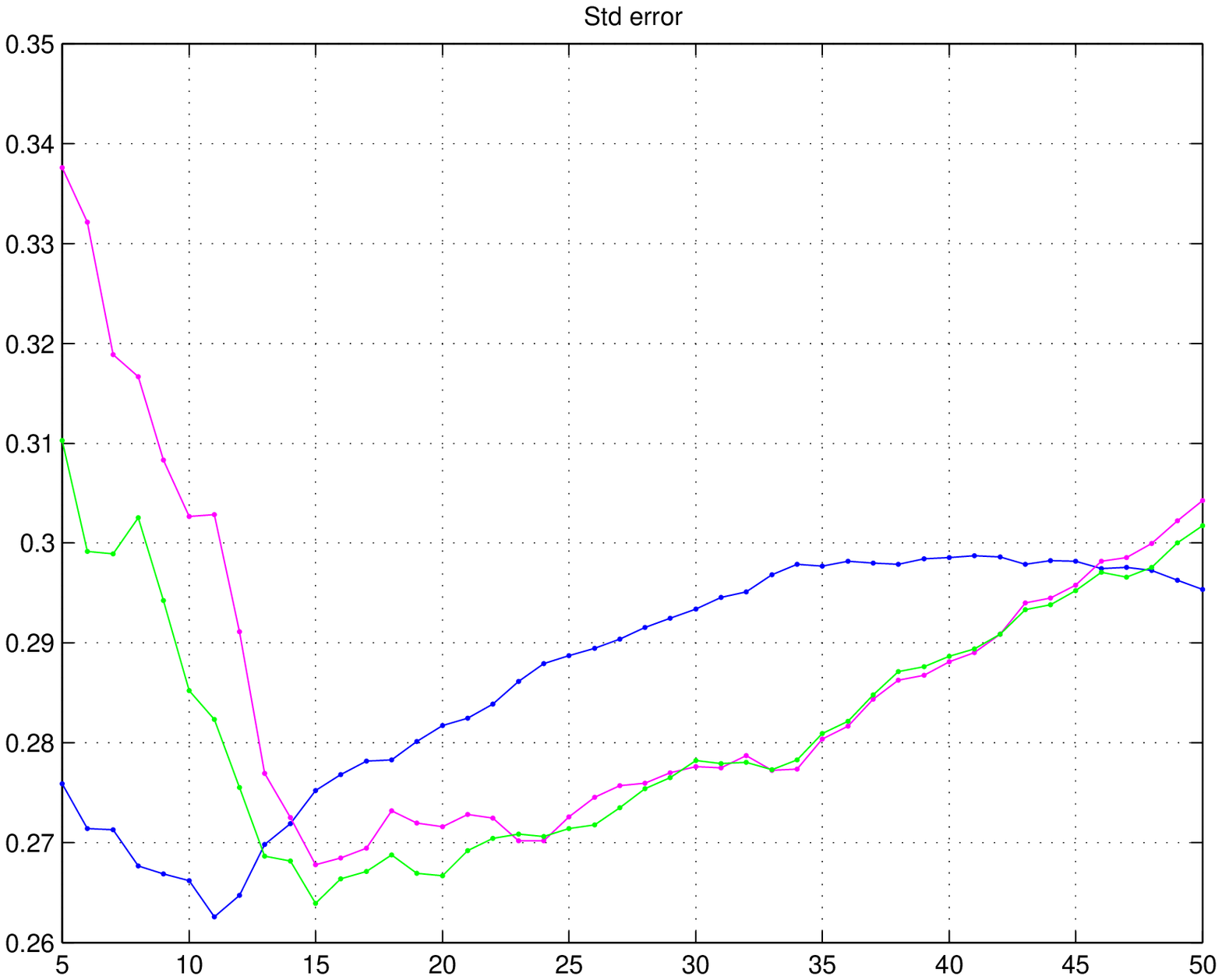}}
  \end{center}
    \caption{
SD of errors for the three predictions described above.
     }
     \label{f26}

\end{figure}

\begin{figure}[!hb]
  \begin{center}
    \scalebox{0.8}{\includegraphics{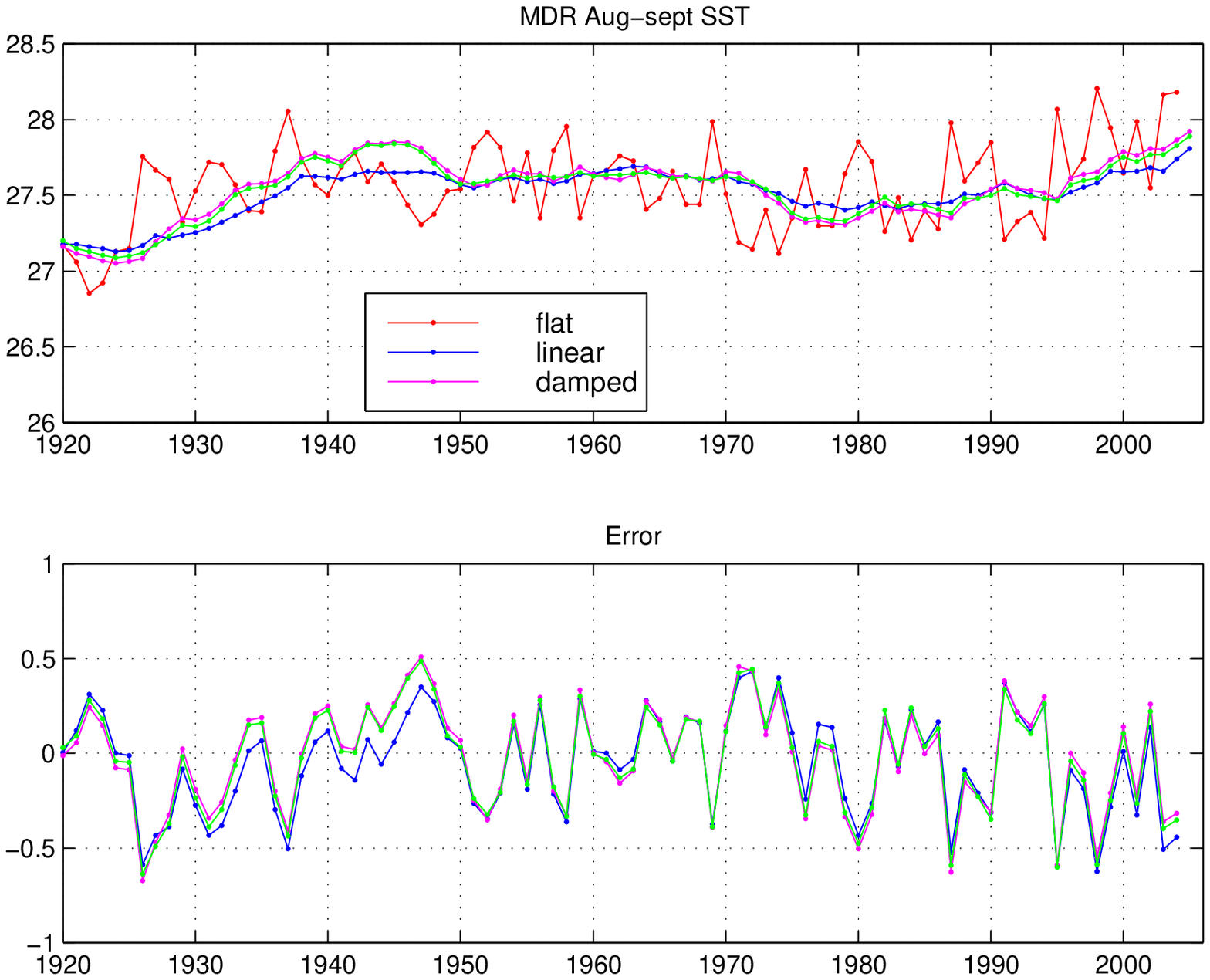}}
  \end{center}
    \caption{
The top panel shows hindcasts for the August-September MDR SST index shown
from the flat-line (blue), best fit linear trend (red) and damped linear trend (green) models,
along with actual values for the index.
The lower panel shows the errors from each of the three predictions.
     }
     \label{f27}
\end{figure}

In this case (see figures~\ref{f24} to~\ref{f27})
the results are not too dissimilar from the June-November results, with the DLT model performing best.

\section{Discussion}

We are interested in the prediction of tropical Atlantic SSTs a year in advance, and have
tested a number of simple statistical prediction schemes on past SSTs to see which would have
given the best predictions. We have considered four SST indices: Gulf and MDR for June-November,
and Gulf and MDR for August-September.

For FL models, the best results come from using either 11 or 12 years of data in all four cases.
For the LT model, the best results come from using between 15 and 26 years of data.
For the DLT model, the best results come from using between 15 and 27 years of data.
When only short periods of data are used (less than 10 years) the FL model always does the best.
When more than 15 years of data are used, the DLT model usually does best.
The best results from the FL and DLT models are typically quite close, and
are typically a little better than the best results from the LT model.

What, then, should we use if we want to predict SSTs for next year?
There is not a great deal to choose between the models. If one wanted to use the same number of years
in all regions, then one might (fairly arbitrarily) choose predictions based on FL with 12 years
or DLT with 15 or 20 years. The DLT method might be preferred since good performance is less sensitive
on the exact number of years chosen than for LT.
Are there any other considerations that should come into play other
than the raw backtesting results? One shortcoming of the whole principle of backtesting is that
it only tells us what might have worked well in the past, and doesn't tell us what will work well in the future.
If the dynamics of SST variability is different now than it was in the past, then backtesting could mislead
us. For this reason, it seems reasonable to choose those methods that rely on less data, since recent
data is presumably more relevant if new processes are occurring. This is another reason to avoid using
the LT model, and might lead us to prefer FL over DLT. Alternatively one
might argue that the recent SST variability may be showing a stronger trend than
on average over the backtesting period. This might then
lead one to choose DLT, which takes the trend into account and so is likely to do well during periods of strong trends.

Overall, we conclude that this study shows that LT should not be used, and that DLT is probably marginally
more useful than FL because of the low sensitivity to the number of years, and the good performance
during periods of strong trends, such as we are now experiencing.


There are a number of areas for future work, apart from the obvious next step of trying to relate SST to hurricane
numbers. One would be to consider statistical methods for combining these different
forecasts. Such combinations may do better than any individual forecasts. Another would be to incorporate
the effects of ENSO, especially on predictions for the MDR region.

\bibliography{meagher1}

\begin{thebibliography}{9}
\providecommand{\natexlab}[1]{#1}
\providecommand{\url}[1]{\texttt{#1}}
\expandafter\ifx\csname urlstyle\endcsname\relax
  \providecommand{\doi}[1]{doi: #1}\else
  \providecommand{\doi}{doi: \begingroup \urlstyle{rm}\Url}\fi

\bibitem[Jewson et~al.(2005)Jewson, Casey, and Penzer]{j90}
S~Jewson, C~Casey, and J~Penzer.
\newblock {Year ahead prediction of US landfalling hurricane numbers: the
  optimal combination of long and short baselines}.
\newblock \emph{arxiv:physics/0512113}, 2005.

\bibitem[Jewson and Penzer(2004)]{j58}
S~Jewson and J~Penzer.
\newblock Optimal year ahead forecasting of temperature in the presence of a
  linear trend, and the pricing of weather derivatives.
\newblock \emph{http://ssrn.com/abstract=563943}, 2004.

\bibitem[Jewson and Penzer(2005)]{j74}
S~Jewson and J~Penzer.
\newblock Weather derivative pricing and the detrending of meteorological data:
  three alternative representations of damped linear detrending.
\newblock \emph{http://ssrn.com/abstract=653241}, 2005.

\bibitem[Khare and Jewson(2005{\natexlab{a}})]{j81}
S~Khare and S~Jewson.
\newblock {Year ahead prediction of US landfalling hurricane numbers}.
\newblock \emph{arxiv:physics/0507165}, 2005{\natexlab{a}}.

\bibitem[Khare and Jewson(2005{\natexlab{b}})]{j88}
S~Khare and S~Jewson.
\newblock {Year ahead prediction of US landfalling hurricane numbers: intense
  hurricanes}.
\newblock \emph{arxiv:physics/0512092}, 2005{\natexlab{b}}.

\bibitem[Klotzbach and Gray(2004)]{klotzbachg04}
P~Klotzbach and W~Gray.
\newblock {Updated 6-11 month prediction of Atlantic basin seasonal hurricane
  activity}.
\newblock \emph{Weather and Forecasting}, 19:\penalty0 917--934, 2004.

\bibitem[Rayner et~al.(2002)Rayner, Parker, Horton, Folland, Alexander, Rowell,
  Kent, and Kaplan]{hadisst}
N~Rayner, D~Parker, E~Horton, C~Folland, L~Alexander, D~Rowell, E~Kent, and
  A~Kaplan.
\newblock {Global analyses of SST, sea ice and night marine air temperature
  since the late nineteenth century}.
\newblock \emph{{Journal of Geophysical Research}}, 108:\penalty0 4407, 2002.

\bibitem[Saunders and Lea(2005)]{saundersl05b}
M~Saunders and A~Lea.
\newblock {Extended Range Forecast for Atlantic Hurricane Activity in 2006}.
\newblock Technical report, {Tropical Storm Risk}, 12 2005.

\bibitem[Sutton and Hodson(2005)]{suttonh}
R~Sutton and D~Hodson.
\newblock {Atlantic ocean forcing of North American and European summer
  climate}.
\newblock \emph{Science}, 309:\penalty0 115--118, 2005.

\end{thebibliography}


\newpage

\newpage

\newpage

\newpage

\newpage

\newpage

\newpage

\newpage

\newpage
\begin{figure}[!hb]

  \begin{center}
    \scalebox{0.8}{\includegraphics{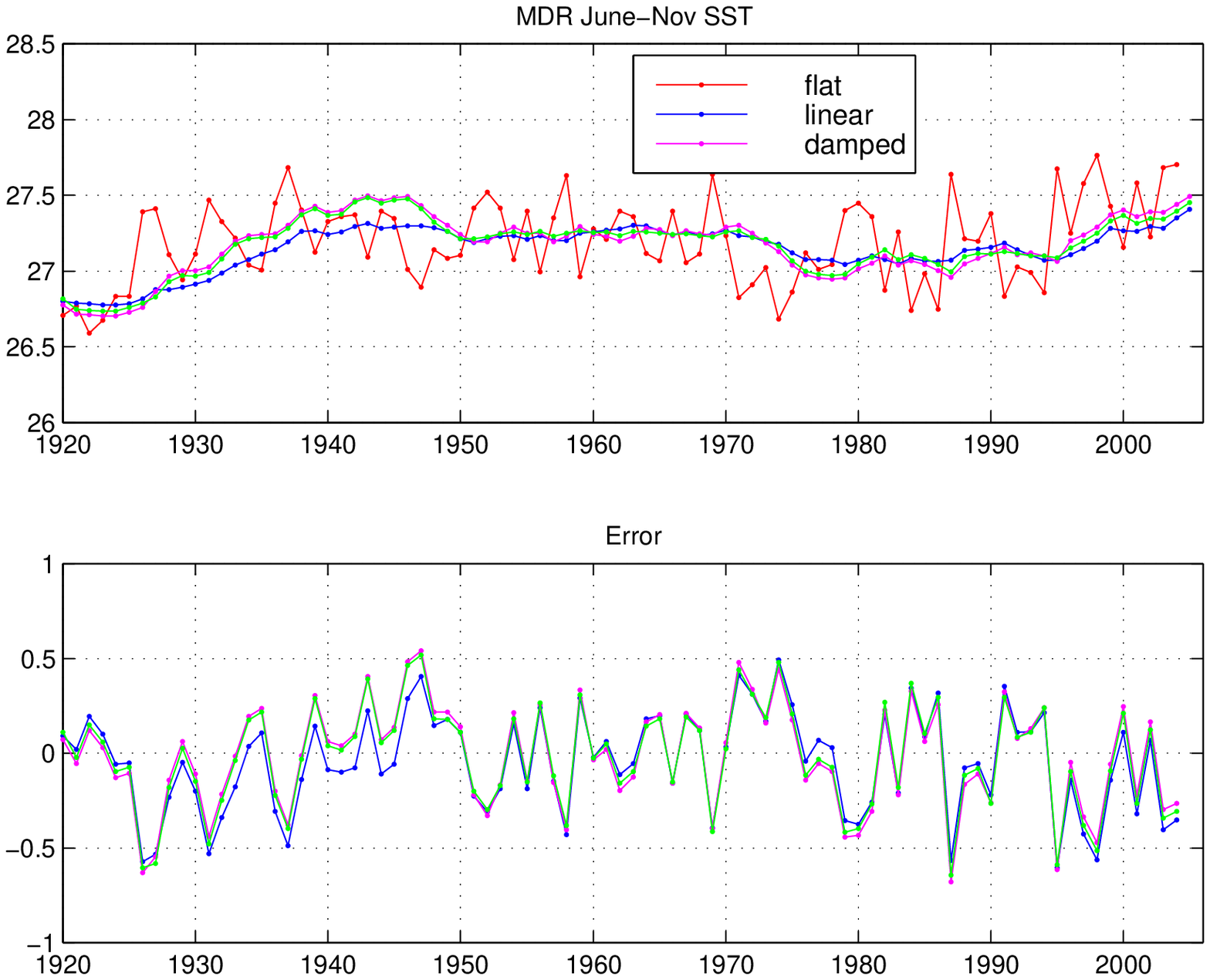}}
  \end{center}
    \caption{
The top panel shows hindcasts for the SST index shown in figure~\ref{f09}
from the flat-line (blue), best fit linear trend (red) and damped linear trend (green) models,
along with actual values for the index.
The lower panel shows the errors from each of the three predictions.
     }
     \label{f19}
\end{figure}

\newpage

\newpage

\newpage

\newpage

\end{document}